\documentclass[prl,showkeys,twocolumn]{revtex4}
\usepackage{color,graphicx,amssymb,amsmath}
\usepackage{hyperref}
\usepackage{natbib} 

\begin{document}

\newcommand{\heading}[1]{{\vspace{0.25truecm}\noindent\textbf{#1.}}}

\title{Unraveling the hidden organisation of urban systems and their mobility flows}

\author{Riccardo Gallotti$^{1}$}
\email{rgallotti@gmail.com}
\author{Giulia Bertagnolli$^{1,2}$}
\author{Manlio De Domenico$^{1}$}
\email{mdedomenico@fbk.eu}

\affiliation{$^{1}$ CoMuNe Lab, Fondazione Bruno Kessler, Via Sommarive 18, 38123 Povo (TN), Italy}
\affiliation{$^2$ Department of Mathematics, University of Trento, Via Sommarive 14, 38123 Povo (TN), Italy}

\begin{abstract}
Increasing evidence suggests that cities are complex systems, with structural and dynamical features responsible for a broad spectrum of emerging phenomena. 
Here we use a unique data set of human flows and couple it with information on the underlying street network to study, simultaneously, the structural and functional organization of 10 world megacities. 
We quantify the efficiency of flow exchange between areas of a city in terms of integration and segregation using well defined measures. Results reveal unexpected complex patterns that shed new light on urban organization. Large cities tend to be more segregated and less integrated, while their overall topological organization resembles that of small world networks. At the same time, the heterogeneity of flows distribution might act as a catalyst for further integrating a city.
Our analysis unravels how human behaviour influences, and is influenced by, the urban environment, suggesting quantitative indicators to control integration and segregation of human flows that can be used, among others, for restriction policies to adopt during emergencies and, as an interesting byproduct, allows us to characterize functional (dis)similarities of different metropolitan areas, countries, and cultures.
\end{abstract}

\keywords{complex networks | integration | segregation | human mobility | urban systems}

\maketitle
 
\section*{Introduction}

Cities are complex systems embedded in the physical space which process information, evolve and adapt to their environment~\cite{barthelemy2019statistical}. To understand how complex systems -- and cities more specifically -- operate, it is thus important to quantify how information is processed in terms of integration and segregation. To this aim, on the one hand many relevant network descriptors have been introduced, based either on topological features or on dynamical ones, or both. On the other hand, integration has been reflected either in how information flow is accounted for by more complex topological models where multiple relationships co-exist simultaneously~\cite{mucha2010community,szell2010multirelational,manlio2013,de2018multilayer}, namely multilayer systems~\cite{kivela2014multilayer,manlio2016}, or in causal effects observed in the time course of systems' units~\cite{schreiber2000measuring,barnett2009granger,runge2012escaping,sugihara2012detecting,stramaglia2014synergy,van2015causal,diez2015information,tononi2016integrated,james2016information,ye2016information}.

Concerning the topological analysis of classical single-layer networks, to date a clear definition of integrated and segregated information flow is still debated and many proxies are used across a broad spectrum of disciplines, ranging from neuroscience to social and urban sciences~\cite{latora2001,newman2004analysis,guimera2005functional,colizza2006detecting,bassett2009human,rubinov2010complex,van2011rich,sporns2013network,centola2015social,deco2015rethinking,cohen2016segregation,aerts2016brain,bertolero2017diverse,bertolero2018mechanistic,yamamoto2018impact,stella2019influence}, often indicating with the same name very different concepts. 

The recent availability of a large amount of human-generated data enables the analysis of urban systems from different perspectives which could not be even considered until a few years ago~\cite{batty2013big}. Consequently, models and analytical tools inspired by complexity science are proliferating. More and more examples are providing convincing evidences of their fruitful application to real cities~\cite{tsai2005quantifying,guerois2008,schwarz2010urban,louail2014mobile,Gately2015,ewing2015compactness}. Applications range from human mobility~\cite{Song2010,louail2015uncovering,Gallotti2016,barbosa2018human} and traffic congestion~\cite{helbing2001traffic,li2015percolation,olak2016,sole2018decongestion, depersin2018global}, to energy consumption~\cite{le2012urban}, air quality~\cite{Stone2008,Uherek2010} and climate~\cite{Martilli2014}, health and well being~\cite{Ewing2014,Newby2014,Rice2015,Li2017}, and the associated topic of accessibility to important facilities like hospitals~\cite{Nicholl2007}. Indeed, the city can be seen as a growing complex system  ~\cite{bettencourt2007growth,bettencourt2013origins} whose spatial organization~\cite{bertaud2004spatial,volpati2018spatial} dynamically experiences a transition from monocentric to polycentric~\cite{louf2013modeling,Louf2014}.

The relative ease of accessing large and detailed data sources describing at the same time the structure and the function of urban systems, puts them in the position of becoming a paradigmatically example over which we can identify the right methodologies allowing us to understand the behaviour of spatially embedded complex systems. A particularly relevant perspective is provided by activity-aware information~\cite{phithakkitnukoon2010activity}, such as the one provided by users of Foursquare -- a leading location intelligence platform -- which allows people to investigate human flows at different scales and thus to reconstruct the functional network of cities with great level of detail~\cite{Noulas2013} and to classify existing activities into a few representative macro-categories (see Methods for details).

In this work, we stratify those human activities to build the functional networks describing the human movements across the urban space of 10 different metropolitan systems spread over three continents. 
To gain novel insights about the functional organisation of the underlying urban ecosystem, we build a multilayer network \cite{manlio2013,manlio2016}, where the flows encode how users move between venues of the same macro-category (e.g., from a pub to another one) and between venues of different macro-categories (e.g., from a pub to a cinema). In the following, we will refer to \emph{intra-layer} flow to indicate movements of the first type, and to \emph{inter-layer} flow to indicate movements of the second type. 

Our main goal is to better characterize the functional organization of a city through the lens of network science. To this aim we measure to which extent different areas of the city facilitate human flows -- i.e., functional integration -- and to which extent there are separate clusters of areas characterized by within-cluster flows larger than between-cluster flows -- i.e., functional segregation -- (see Methods for details)~\cite{bullmore2012economy}. By considering those measures simultaneously, it is possible to characterize how well human flows mix through the city according to the existing distribution of venues and the way residents use them. In fact, the dichotomy between integration and segregation -- often improperly used as antonyms -- is relevant for improving our understanding of the interplay between the urban structure, social relationships and human behavior.

At the same time, to investigate the coupling between the structure of a city and the dynamics of its inhabitants, we also study the integration and segregation of the structural networks of these cities reconstructed from Open Street Map~\cite{boeing2017osmnx}. See Fig~~\ref{fig:first_figure} and Methods for more details on the definition of the structural and functional networks.

\section*{Results}

\begin{figure*}[!t]
\begin{center}
\includegraphics[angle=0,width=\textwidth]{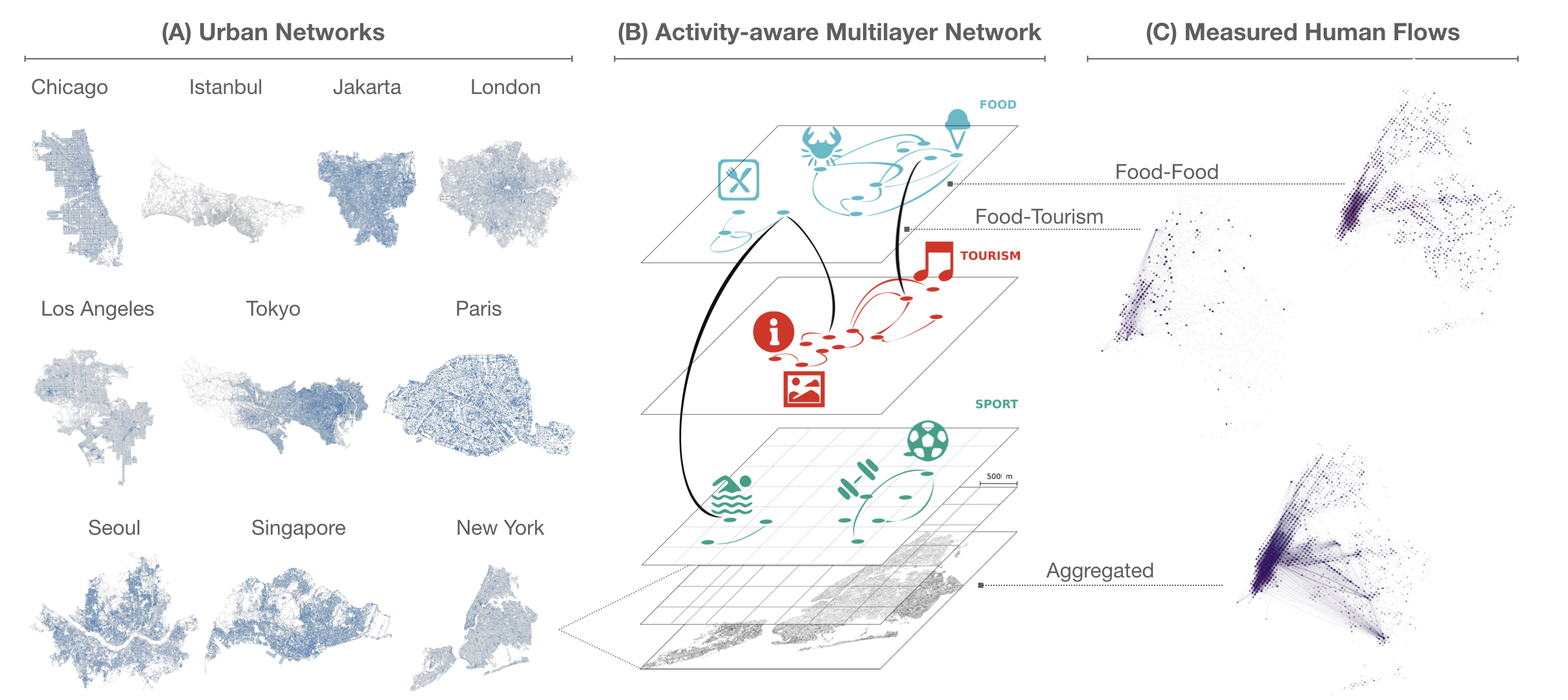}
\end{center}
\caption{
\textbf{Modeling Structure and Function of Urban Systems.}  \textbf{Left:} Urban structural backbone of the 10 megacities considered here, as described from their street networks (data obtained from Open Street Map~\cite{boeing2017osmnx}). \textbf{Middle:} Urban functional networks described by the Foursquare data. The nodes are obtained by dividing the area analysed into cells of 500m~$\times$~500m. The edges are subsequent check-ins that might be between activities of the same type (intra-links: e.g. Food-Food, Tourism-Tourism) or different types (inter-links: e.g. Food-Tourism, Food-Sport). The collection of layers and inter-layer flows defines a multilayer network~\cite{manlio2013,kivela2014multilayer,manlio2016}, i.e., a multidimensional functional representation of the urban areas. \textbf{Right:} The mobility flows between areas are captured as the edges' weights. In the example, describing New York City, we can observe the different spatial distribution of flows between and across different activity layers (see also Fig.~\ref{fig5}a).
\label{fig:first_figure}
}
\end{figure*}

\heading{Overview of the data sets} 
The Foursquare data made available for the Future Cities Challenge~\cite{FCC} describe 24 months of check-ins between April 2017 and March 2019 (included). The 10 world mega-cities included in the challenge are Chicago, Istanbul, Jakarta, London, Los Angeles, Tokyo, Paris, Seoul, Singapore and New York City (represented as example in Fig.~\ref{fig:first_figure} right). The extensive characteristics of the datasets are shown in Tab.~\ref{fig:table_data}. The flows between different areas are derived by subsequent anonymized check-ins to the Foursquare's location-based services and coarse grained with a 500m $\times$ 500m granularity (see Fig.~\ref{fig:first_figure} middle, and Methods). In the data provided, check-ins are already aggregated by couple of venues (origin and destination), month, and hour of the day (morning, midday, afternoon, night, and overnight). 

The Open Street Map data has been obtained using the OSMNX python library~\cite{boeing2017osmnx} (see Fig.~\ref{fig:first_figure} left). The urban area selected has been set to matches the cells covered by the Foursquare venues. The structural network has been reduced to a lattice-like form of the same granularity as the urban flow, so that all nodes in the structural network find their correspondence in the functional network. Differently from the functional one, the structural network is purely topological, as an undirected link between two cells exists if at least one street connects the two areas. 

\heading{Quantifying Integration and Segregation} 
As previously mentioned, we characterize the organization of the city through measures of integration and segregation. To avoid confusion in the reader, it is worth remarking that our measures of integration and segregation are those established in the field of network neuroscience~\cite{cohen2016segregation}, rather than being associated to the traditional social concepts, and are thus not related to population or cultural mixing~\cite{louf2016patterns}, but only to how cities are lived by their users.
Integration quantifies, in terms of information exchange efficiency, the ability of a city to favor the flow of people across its areas, and is measured by means of the global communication efficiency GCE, specifically normalized to correctly compare the efficiencies of weighted and un-weighted networks~\cite{bertagnolli2020quantifying}.
Segregation, on the other hand, evaluates the strength of segregated communities, areas of the city with strong flows inside the area and weak inter-areas flows and is estimated as the maximal modularity $Q^\ast$~\cite{Newman2004b} of the network (see Methods for further details).

\heading{Structural vs Functional Networks} 
Having identified two measures suitable for comparing different cities and types of networks, we begin our analysis by mapping the link between integration and segregation in both the structural road networks and the single layer flow networks, obtained aggregating for each city inter-layer and intra-layer flows over the whole temporal extension of the dataset, which describe the functional use of the city by individuals.

\begin{figure*}[!t]
\begin{center}
\begin{tabular}{cc}
\raisebox{2.5cm}{(a)} \includegraphics[angle=0, width=0.45\textwidth]{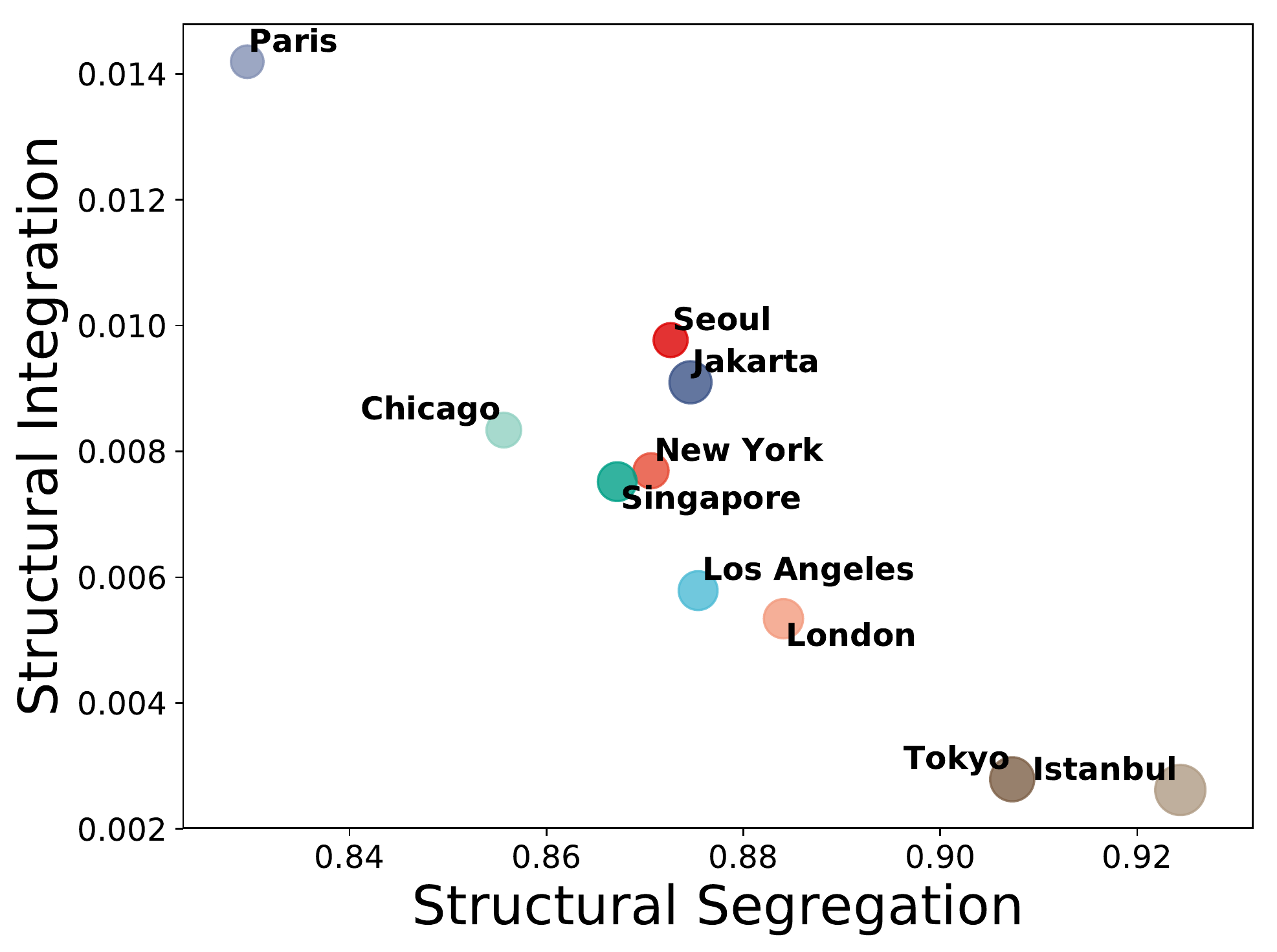}&
\raisebox{2.5cm}{(b)} \includegraphics[angle=0, width=0.44\textwidth]{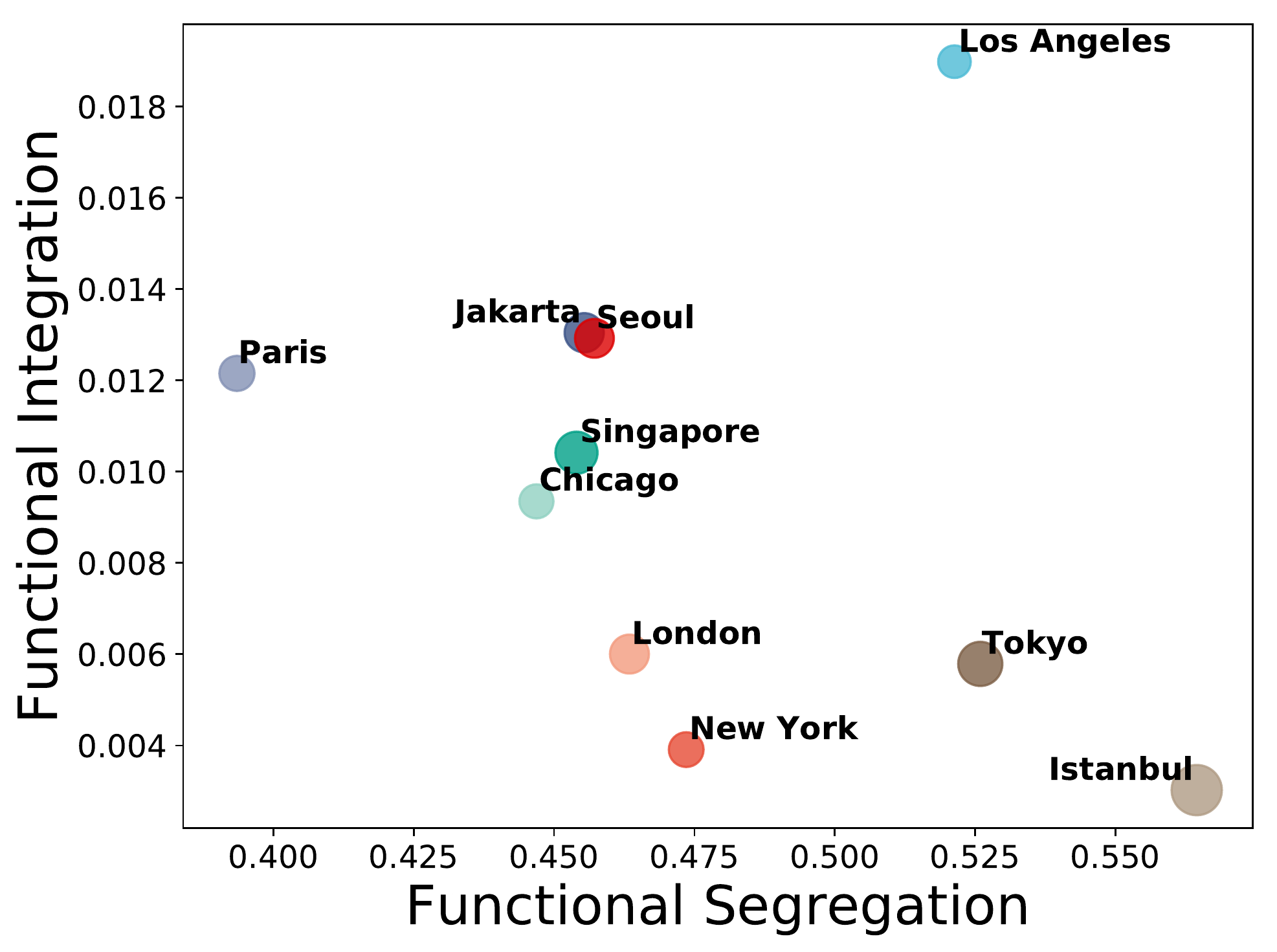} \\
\raisebox{2.5cm}{(c)} \includegraphics[angle=0, width=0.44\textwidth]{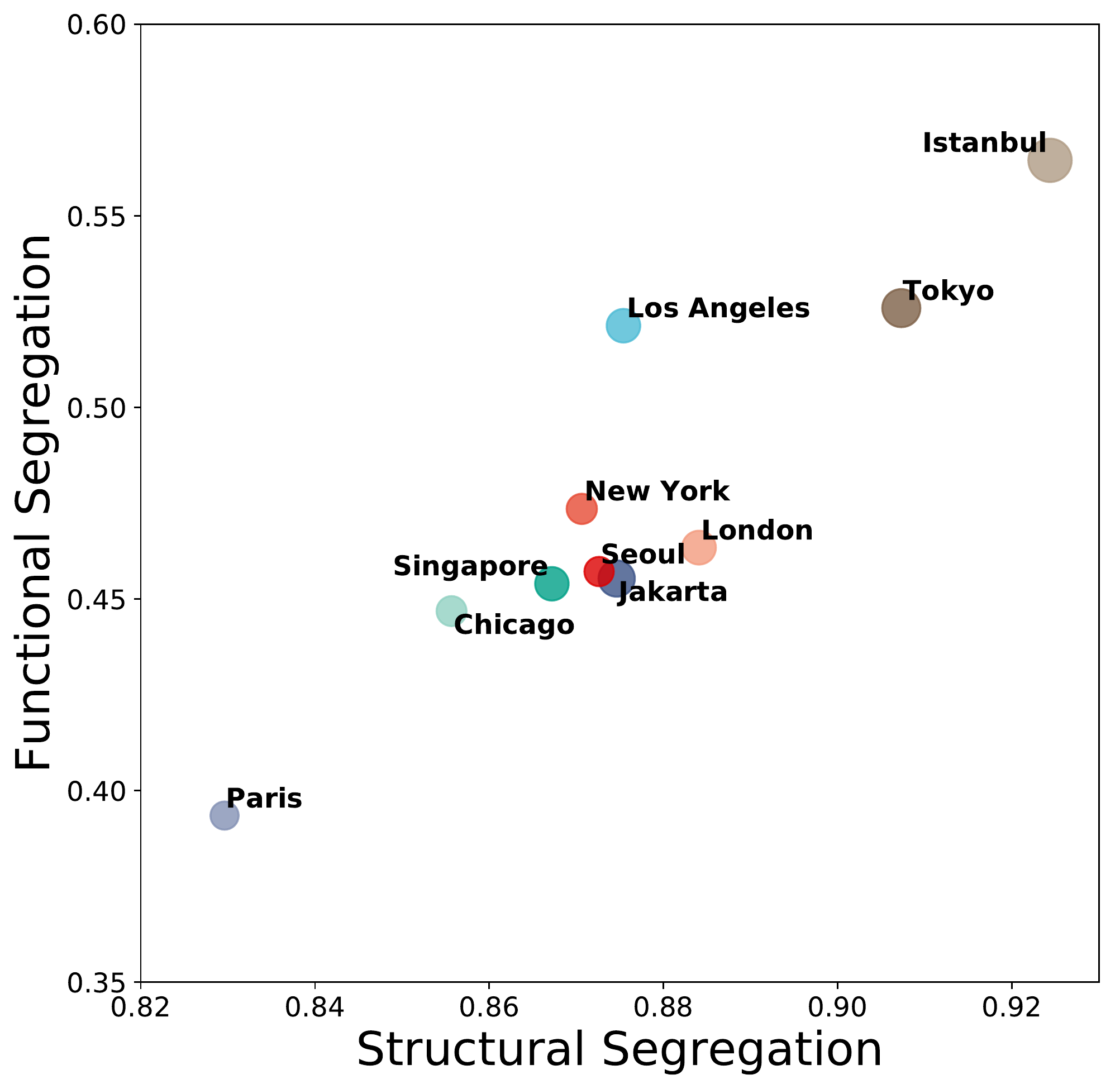} &
\raisebox{2.5cm}{(d)} \includegraphics[angle=0, width=0.45\textwidth]{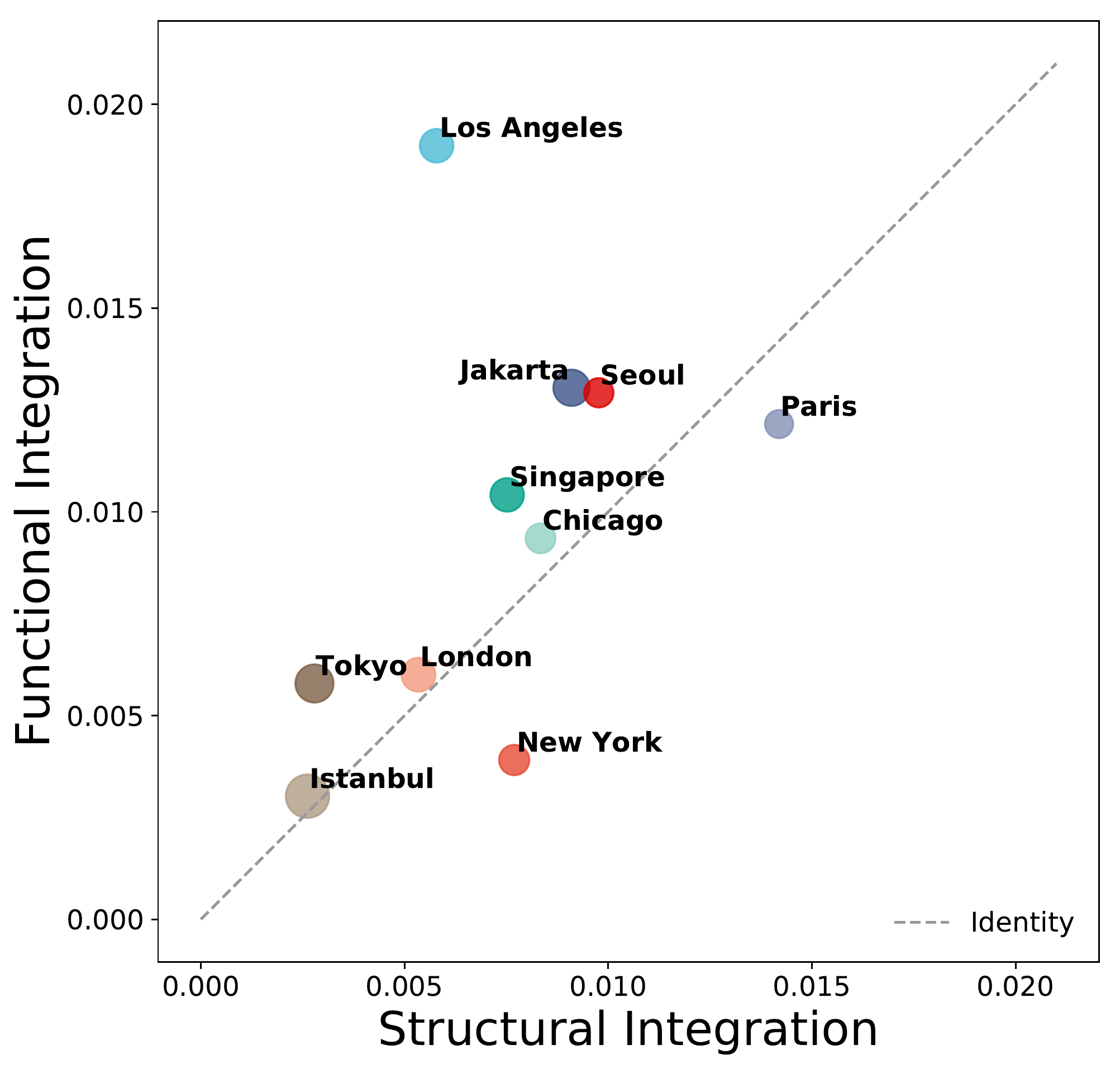}\\
\end{tabular}
\end{center}
\caption{
{\bf Structural vs Functional organization of cities measured by means of Segregation and Integration.} 
{\bf (a)} Structural Integration vs Segregation. Analyzing the measures of segregation ($Q^\ast$) and Integration (GCE) for the topological un-directed network describing the road structure of cities we observe a very strong anti-correlation (Pearson $r=-0.92$).
{\bf (b)} Functional Integration vs Segregation. The same measures of the weighted network describing the mobility flows display clear deviations from the anti-correlation of integration and segregation, in particular for the city of Los Angeles.
{\bf (c)} Structural vs Functional Segregation. The measures of segregation for the two types of networks are strongly correlated (Pearson $r=0.91$) but differ in value.
{\bf (d) }Structural vs Functional Integration. The measures of integration for the two types of networks deviate from perfect correlation (again due to the deviation of Los Angeles) but are very similar in value.
In all panels, the dimensions of the circle is proportional to the size of the area considered.
}
\label{fig2_str_int}
\end{figure*}

The results, displayed in panels (a) and (b) of Fig.~\ref{fig2_str_int}, suggest that, in general, higher values of segregation are associated to lower values of integration, as common sense would suggest. However, we also observe clear deviations from this trend, the major one being the functional network for the city of Los Angeles appearing to be much more integrated than what would be expected by its relatively high level of segregation.

Of particular interest is the comparison of structural and functional properties of the same systems (panels (c) and (d) of Fig.~\ref{fig2_str_int}). The segregation, estimated through the lens of modularity, seems to systematically deviate, with the functional flow network being less segregated than the structural network even if the values for the different cities are highly correlated. The integration instead, studied with an indicator specifically developed for allowing this type of comparisons~\cite{bertagnolli2020quantifying} corresponds also numerically for the very different structural and functional network, and this perfect correspondence reveals a divergence between structural and functional properties of the city of Los Angeles.

\heading{What determines integration and segregation}
In order to understand what lies behind the pattern of anti-correlation between integration and segregation observed in Fig.~\ref{fig2_str_int}, we generate spatially embedded networks that attempt at reproducing the key feature of the urban functional networks using two widely used null models: i) the Watts-Strograts (WS) small world networks obtained through rewiring of a regular lattice; ii) the Random Geometric Networks (RGN) obtained by linking two randomly placed points if their distance falls below a fixed threshold $r$ (see Methods). Also for the RGNs we proceeded with random rewiring and, in both cases, the probability of rewiring is indicated by $p$.

\begin{figure*}[!t]
\centering
\includegraphics[angle=0,width=0.66\textwidth]{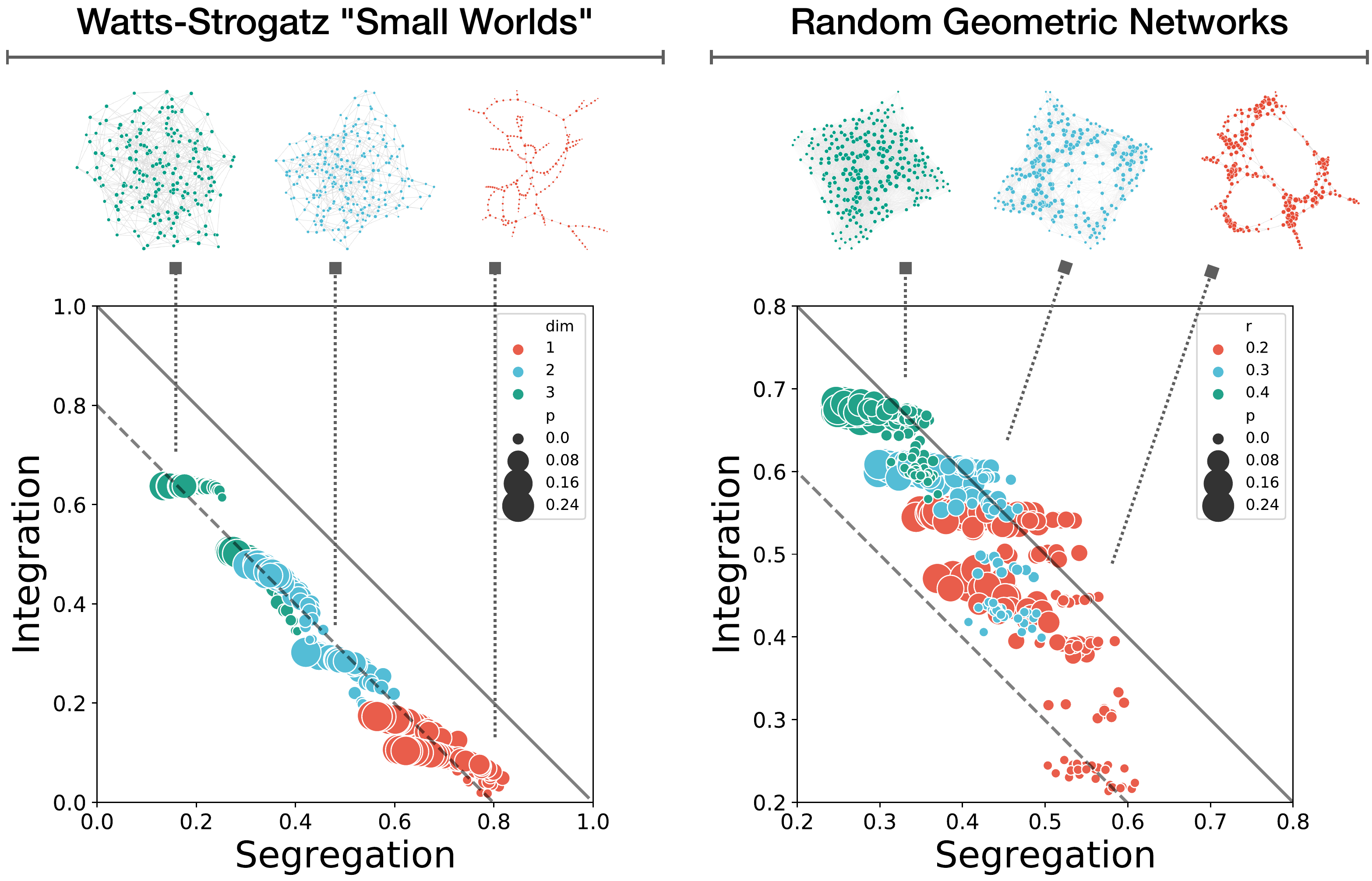}
\includegraphics[angle=0, width=0.305\textwidth]{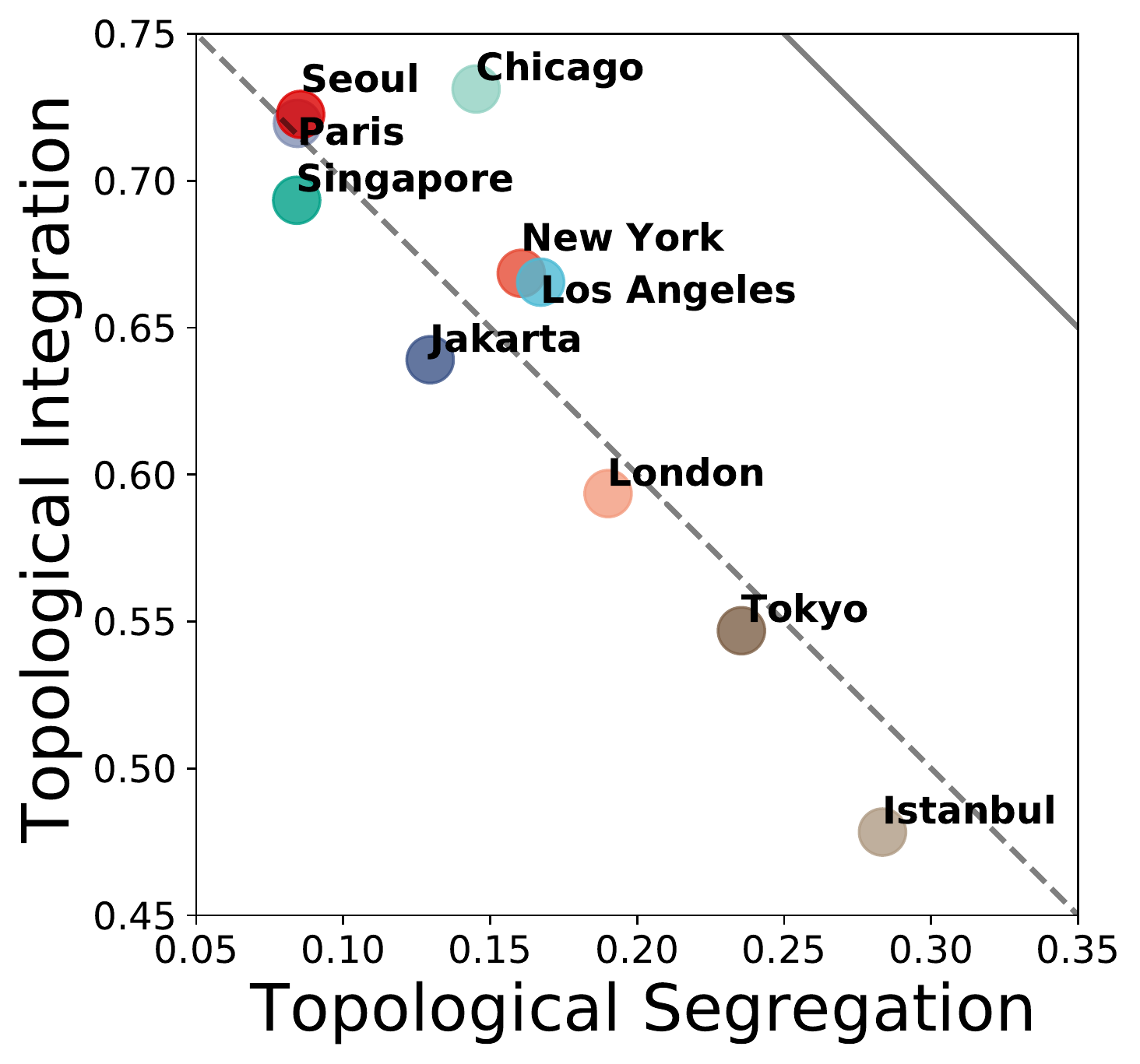}
\caption{\label{fig:synth} 
\textbf{Simulating the functional organization of synthetic urban models.} 
\textbf{Left:} Small-world networks according to the Watts-Strogatz model (see Methods) with different rewiring probabilities (encoded by size) and dimensions (from 1D to 3D, encoded by color). \textbf{Middle:} Random Geometric Networks (see Methods) with different characteristic spatial scale (encoded by color) and different rewiring probabilities (encoded by size). Clusters here fall above what observed for WS model.
\textbf{Right:} The functional organization of real cities, observed thorough the lens of the topological networks derived from the Foursquare flows (see Methods), follow the same trend as in the that of WS networks. 
In all panels, the dashed line represents the linear regression relating integration and segregation for the WS model, whereas the solid line is $y=1-x$ and it is shown as a reference.
}
\end{figure*}

In Fig.~\ref{fig:synth} we observe that for both null models we reproduce the same anti-correlation pattern observed for real networks, but also see that rewiring is strongly reducing segregation and increasing integration in a way that breaks the linear relationship between the two quantities. Moreover, since by generating them we can control all features of the WS and RGN networks considered, we are able to isolate the leading factors behind this pattern. For WS, integration grows and segregation drops as the network dimensionality grows. The same happens for RGN as the radius $r$ grows. Indeed, both increased dimensionality and $r$ leads to generating networks with a higher edge density,  allowing us to isolate the important role played by edge density in dictating the state of integration and segregation of spatial networks. For topological (i.e. not weighted) networks the Global Communication Efficiency, used to estimate integration, grows as the edge density grows. This is indeed what we observe in Supplementary Fig.~\ref{SI:GCE_ed}, while a less tight correlation can be observed for segregation in Supplementary Fig.~\ref{SI:segregation}. 

However, the values observed in Fig.~\ref{fig2_str_int} (b) deviate sensibly by those describing the networks we generated in Fig~\ref{fig:synth}. This because the urban functional networks are defined as weighted networks, while our null models do not describe weights. Indeed, if we reduce the urban functional networks to a purely topological undirected network, we see in Fig.~\ref{fig:synth} (right) that the numerical values of topological urban functional networks correspond to those described by WS model (dashed line). 

\begin{figure*}[!t]
\begin{center}
\begin{tabular}{cc}
\raisebox{2.5cm}{(a)} \includegraphics[angle=0, width=0.45\textwidth]{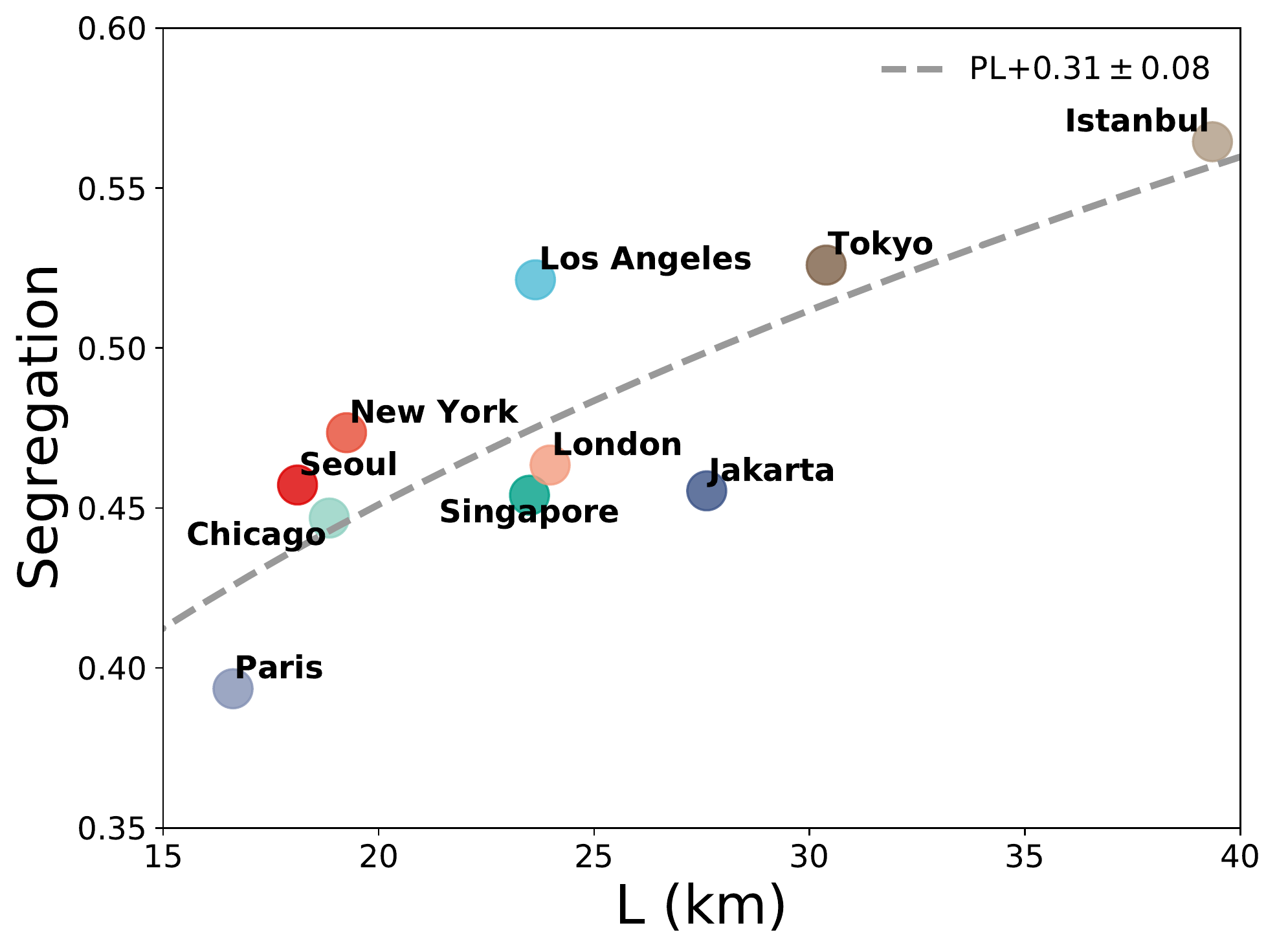} &
\raisebox{2.5cm}{(b)} \includegraphics[angle=0, width=0.45\textwidth]{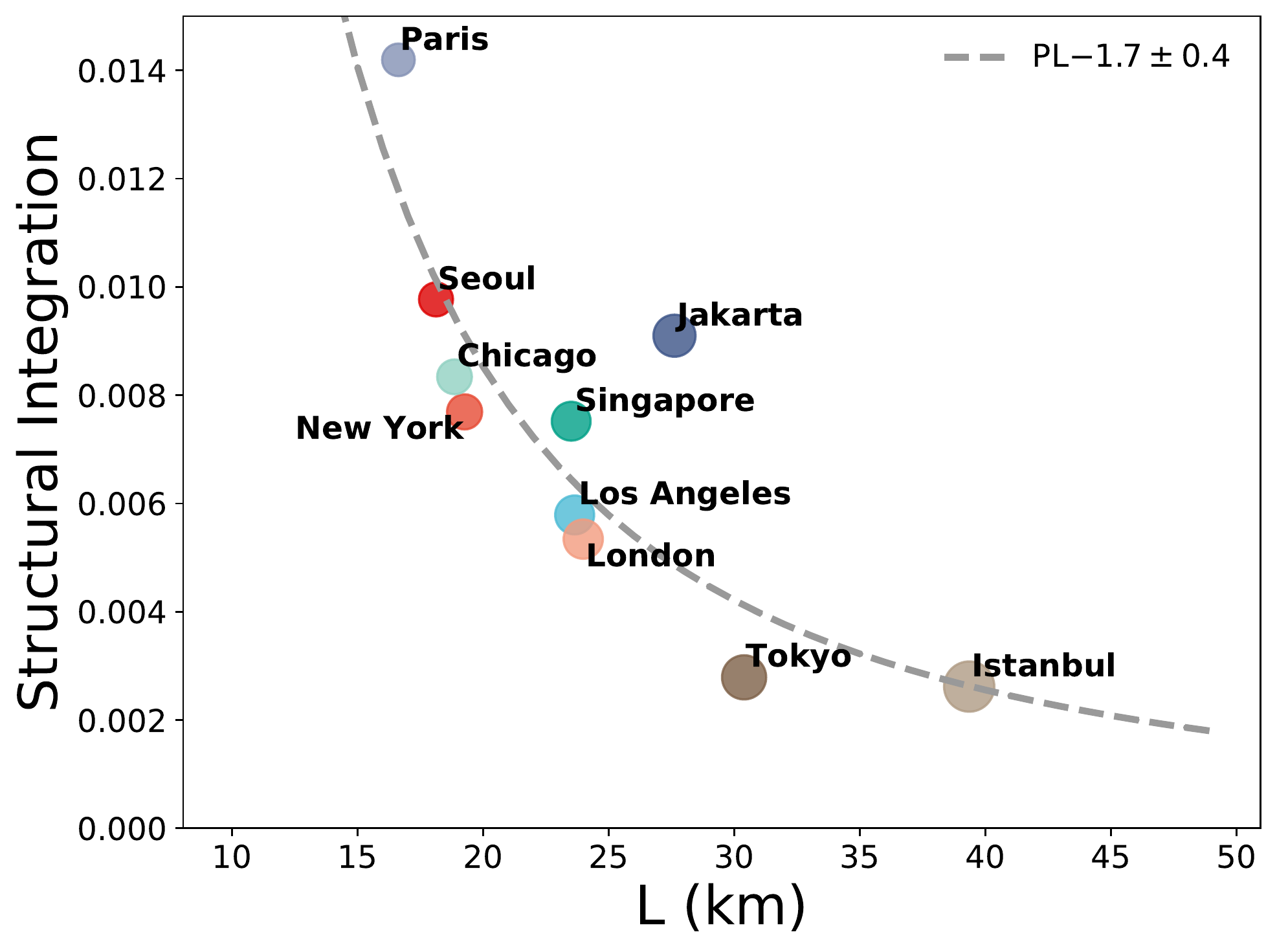}  \\
\raisebox{2.5cm}{(c)} \includegraphics[angle=0, width=0.45\textwidth]{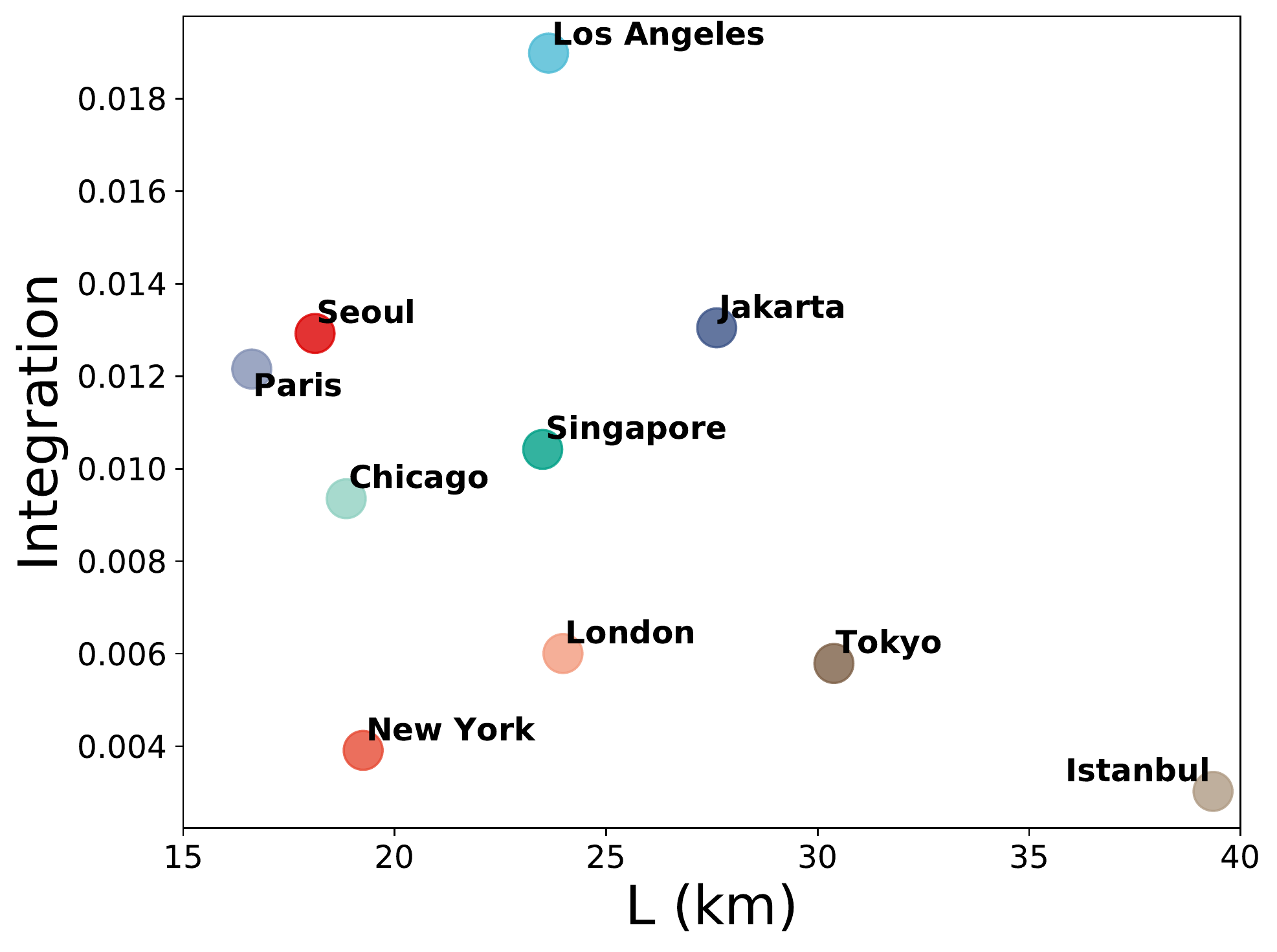}  &
\raisebox{2.5cm}{(d)} \includegraphics[angle=0, width=0.45\textwidth]{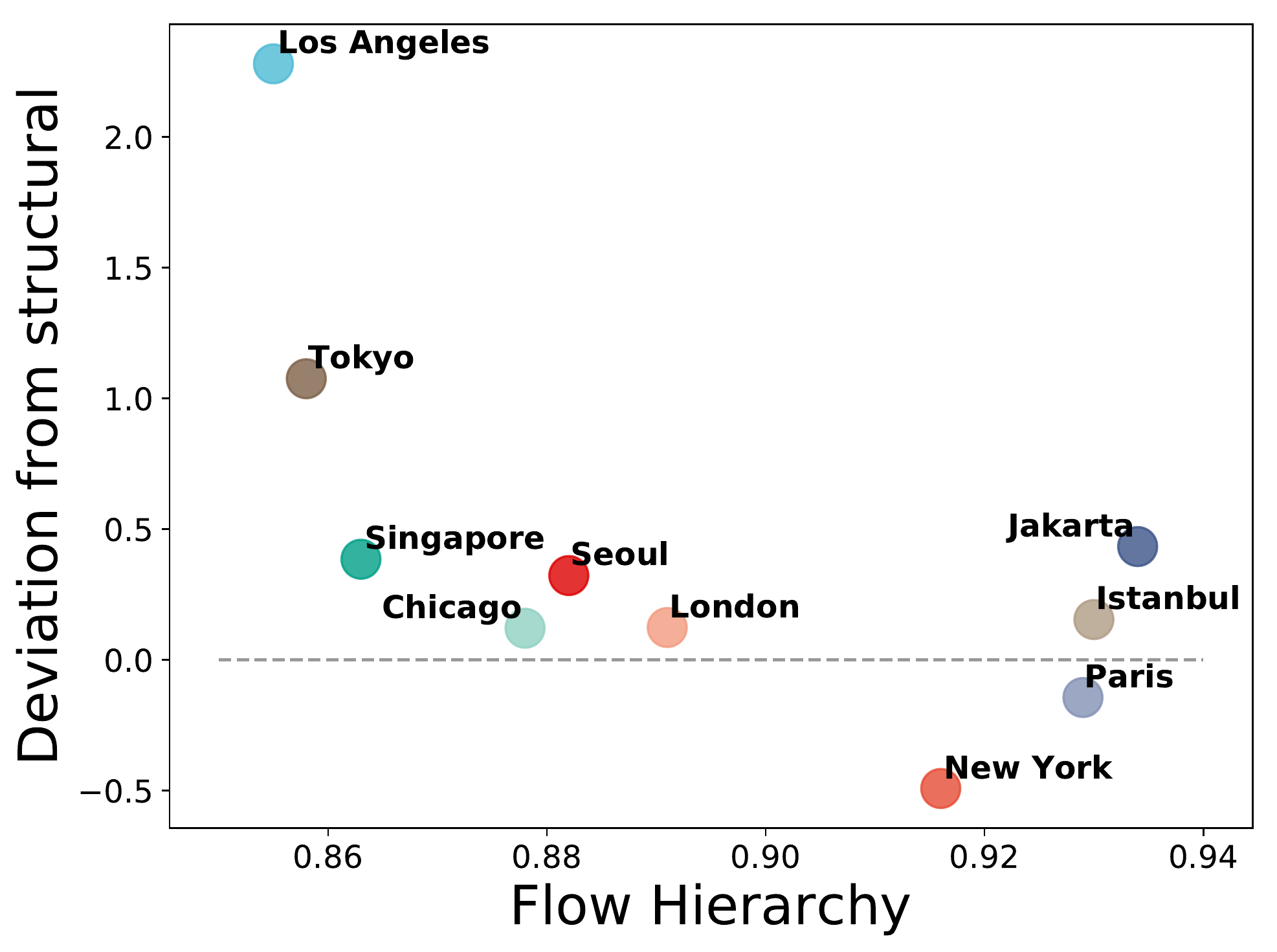} \\
\end{tabular}
\end{center}
\caption{
{\bf Understanding Functional Segregation and Integration.} 
While the functional segregation {\bf (a)} and structural integration {\bf (b)} show a clear dependency over city size, functional integration {\bf (c)} is not simply determined by how big is a city. 
In {\bf (d)}, we plot the deviation between functional and structural integration, computed as $(GCE_{funct}-GCE_{struc})/GCE_{struct}$ vs the values of flow hierarchy for the same cities computed in \cite{bassolas2019hierarchical} from another dataset. 
}
\label{fig_4}
\end{figure*}

To isolate the driving factors determining a city integration and segregation we have to expand from the ideal world of synthetic models and find instead guidance from the methods commonly adopted to investigate the physics of cities. Many properties of cities are known to be power law functions of population size~\cite{bettencourt2007growth}. Here, we are not in the position of deriving with precision the population in the area defined by the Foursquare data, and we use instead as measure of the city size the square root of the area covered ($L= \sqrt{A}$) which is also a proxy for the average length of a trip in a city \cite{louf2013modeling}. We therefore plot in Fig. \ref{fig_4} (a,b,c) the values of Functional Segregation and Structural and Functional integration against $L$ (see Supplementary Figure~\ref{SI:city_size} to see how other network indicators scale). In our case, the sizes of the cities considered are not diverse enough for initiating a meaningful discussion based on the value of the exponents observed (that are reported in panels (a) and (b) only to support future studies on the matter). We focus indeed on the fact that a power law scaling is able to explain most of the variance observed for Functional Segregation ($R^2=0.67$) and Structural Integration ($R^2=0.71$) but totally fails at predicting the values of Functional Integration ($R^2=0.05$). In other words, size matters. In particular it matters for functional segregation, also linked to the total flow circulating over the network (Supplementary Fig.~\ref{SI:segregation} (c)): in fact, as observed in ~\cite{gallotti2014anatomy}, it can be expected to grow proportionally with population. However, there is something more that is strongly influencing functional integration and makes it deviate from the structural integration (as seen in Fig.~\ref{fig2_str_int} d)). This extra factor is determined by how flows are distributed in the network. To show this, in Fig.~\ref{fig_4} (d) we compute how much the weighted functional networks deviate from the values estimated from the structural network as $(GCE_{funct}-GCE_{struc})/GCE_{struct}$, and plot it against the flow hierarchy estimated for the same city from another dataset (numerical values available in~\cite{bassolas2019hierarchical}). A low flow hierarchy indicates that larger fraction of movements are expected to be between strong mobility hubs and less active areas. This means that, in general, excess of integration is expected when marginal areas are more strongly connected. This appears similar to what observed in hierarchical modular brain networks, which are locally segregated, but global neuronal operation integrate segregated functions~\cite{park2013structural}.

Lastly, using the RGN model we also measured the importance of the spatial extension of the network. Fixing the radius below which nodes are connected, we find (see Supplementary Fig.\ref{SI:RGG_scaled_QvsGCE}) that the largest the area ($A= L^2)$ covered by a square RGN the more the network is segregated and the less it is, at the same time, integrated. Indeed, here again integration and segregation seem to be very strongly correlated and increasing the radius have a similar effect as reducing the spatial extension.

\heading{Cities within a city}

Having understood the behaviours of integration and segregation of cities at an aggregated level, is worth checking if this pattern is an intrinsic feature of urban systems or if it is proper of some specific activity layers. Indeed, the metadata of the venues include a \emph{category} field which describes the type of venue in great detail (e.g.:  Knitting Stores, Mini Golf Courses, Rock Clubs, \dots). We defined a set of macro-categories we used to aggregate categories in limited number of layers (see Methods and Fig.~\ref{fig:first_figure} middle). 

\begin{figure*}[!t]
\begin{center}
\raisebox{2.5cm}{(a)} \includegraphics[angle=0, width=0.8\textwidth]{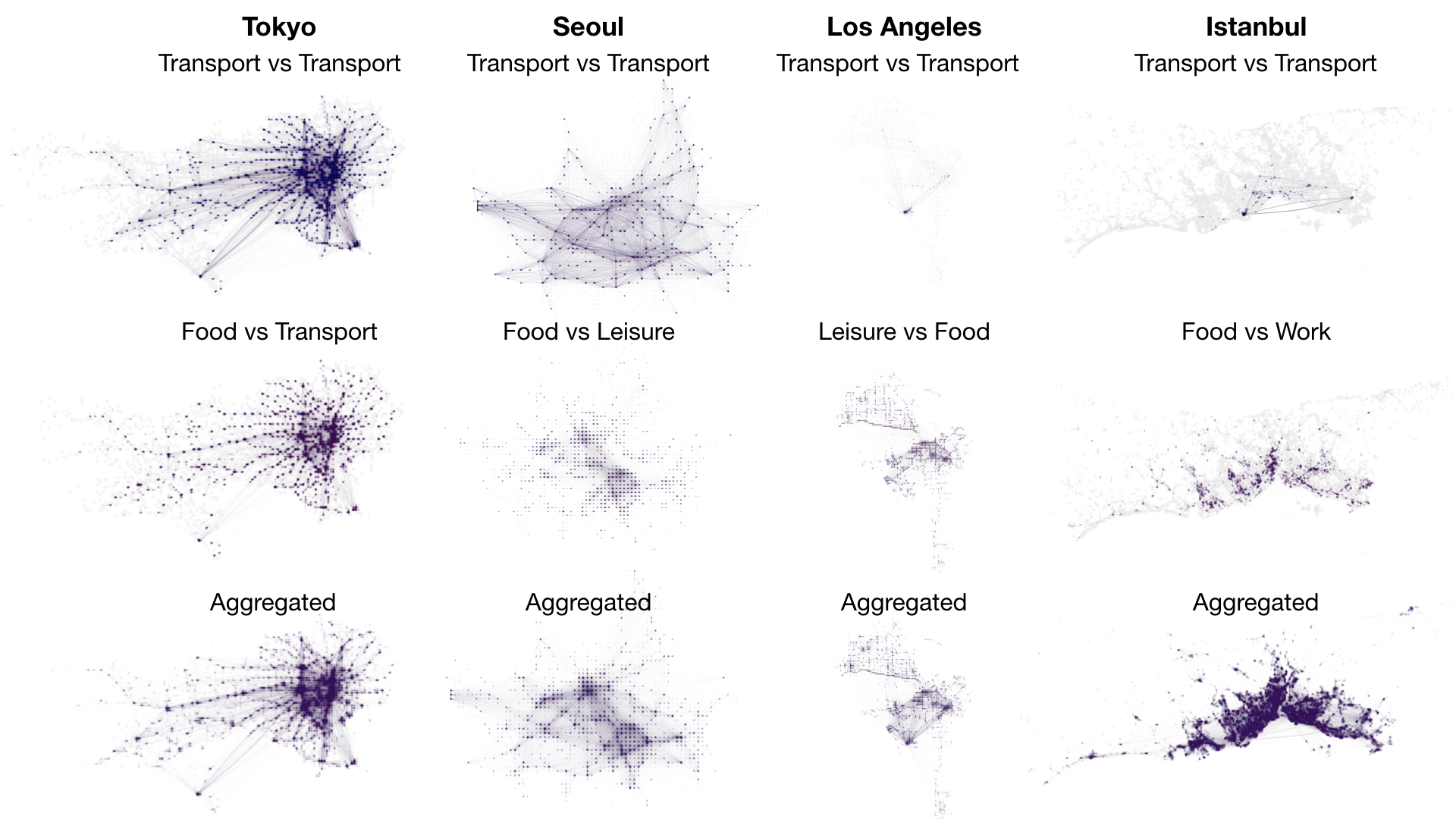}
\begin{tabular}{cc}
\raisebox{2.5cm}{(b)} \includegraphics[angle=0, width=0.4\textwidth]{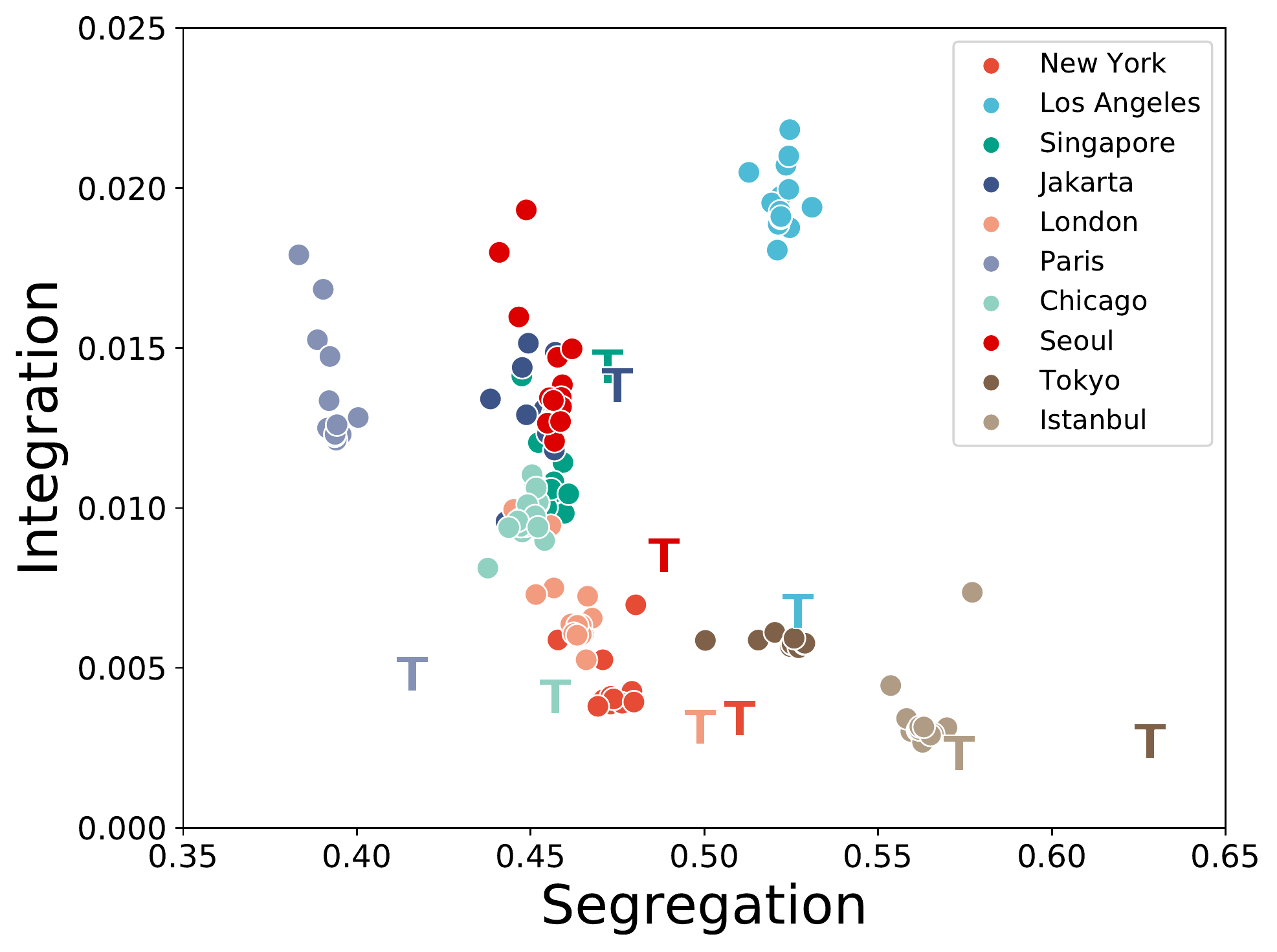} &
\raisebox{2.5cm}{(c)} \includegraphics[angle=0, width=0.4\textwidth]{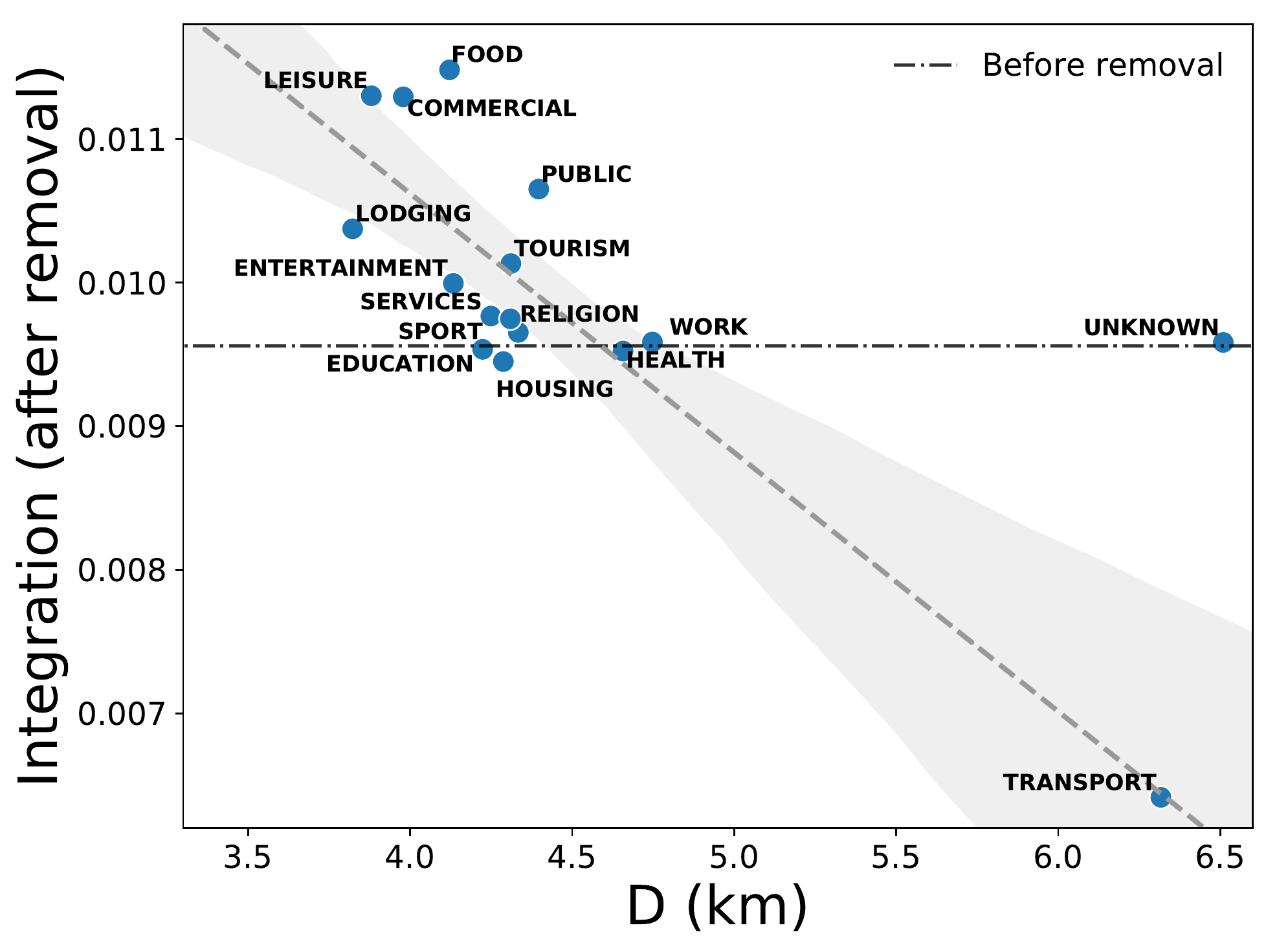} \\
\end{tabular}
\end{center}
\caption{
\textbf{Disentangling functional flows.} 
\textbf{(a)} We illustrate the strikingly distinct views on the functional organization of a city extracted by isolating intra- or inter-layer flows. These maps outline the different ``cities within the city'' which we isolate by decoupling the urban flows into activity-aware multilayer networks.
\textbf{(b)} We define the multilayer networks of human flows for each city (encoded by color) by stratifying flows according to different macro-categories used in this work (see Methods).
Each point corresponds to integration and segregation measured after removing a specific layer of activities. The letter `T' marks values associated to the removal of the transport layer, which strongly influence a the urban functional connectivity (see Fig.\ref{fig_6}). 
\textbf{(c)} Average functional integration for different activity categories. We observe a relationship between the average distance covered $D$ in movement inside one layer and the
value of integration (see Supplementary Figure \ref{SI:Q_removal_D} for segregation). The regression is done excluding the outlier the unclassified venues  ``unknown'' which removal appears not to influence a city's functional integration.
}
\label{fig5} 
\end{figure*}

In Figure~\ref{fig5} (a) we can visually inspect some examples of activity-aware layers. Remarkably, for all the cities considered in this study, the intra-layer connectivity characterizing the transport layer provides a natural link between our functional analysis and the underlying structure of the city. In the data, however, it can be clearly seen in cities where public transport is well developed and largely used, such as Tokyo or Seoul, way more than cities where private transportation is dominant, such as Los Angeles and Istanbul. 

By disentangling the mobility flows into a multilayer network structure (see Methods and Fig.~\ref{fig:first_figure} right), we are able to quantify the differences in the functional organization of human flows between different types of activities or different hours of the day of month of the year (See Supplementary Figures ~\ref{SI:hours} and~\ref{SI:months}), enabling the identification of different ``cities within the city'' which indeed shows clear dissimilarities in terms of both functional integration and segregation.

To this aim, we perform targeted attacks on each layer of the corresponding multilayer network and measure the response of the systems in terms of changes in segregation and integration. In Fig.~\ref{fig5} (b) we observe how removing those flows coming from a specific activity type significantly changes urban functional segregation and integration. 
This is especially true if the activity is Transport, whose removal yields the rightmost outliers in the figure. An even stronger variation is observed in the integration and segregation restricted to movements between similar layers (See Supplementary Fig.~\ref{SI:single_layer}).

To better understand these differences, in Fig.~\ref{fig5} (c) we link the average values of integration measured for flows between the same categories across all cities with the corresponding weighted average of geographical distances between nodes. We observe a bulk of correlated points and two outliers: one the natural long-range linking layer of transportation, the other the locations not associated to a macro category and left as ``unknown'' (see Methods). Excluding ``unknown'' that does not seem to influence integration at all, we observe a clear effect: removing the transport layer strongly disrupts integration, while removing short range layers actually improves it. In Supplementary Fig.~\ref{SI:Q_removal_D} we could conversely see how, again with the notable exception of the removal of the Transport layer, the segregation of cities remains relatively unchanged after single layer removal.
The results of this analysis points out that is possible to close restaurants, leisure and commercial activities while keeping a city functional and, possibly, even more integrated. This perspective provides new insight on the effects of restriction policies adopted during emergencies by quantifying a hidden, systemic, social costs and benefits associated to the closure of different kind of activities in time of a pandemic emergency.

\begin{figure*}[!t]
\begin{center}
\begin{tabular}{cc}
\raisebox{2.5cm}{(a)} \includegraphics[angle=0, width=0.45\textwidth]{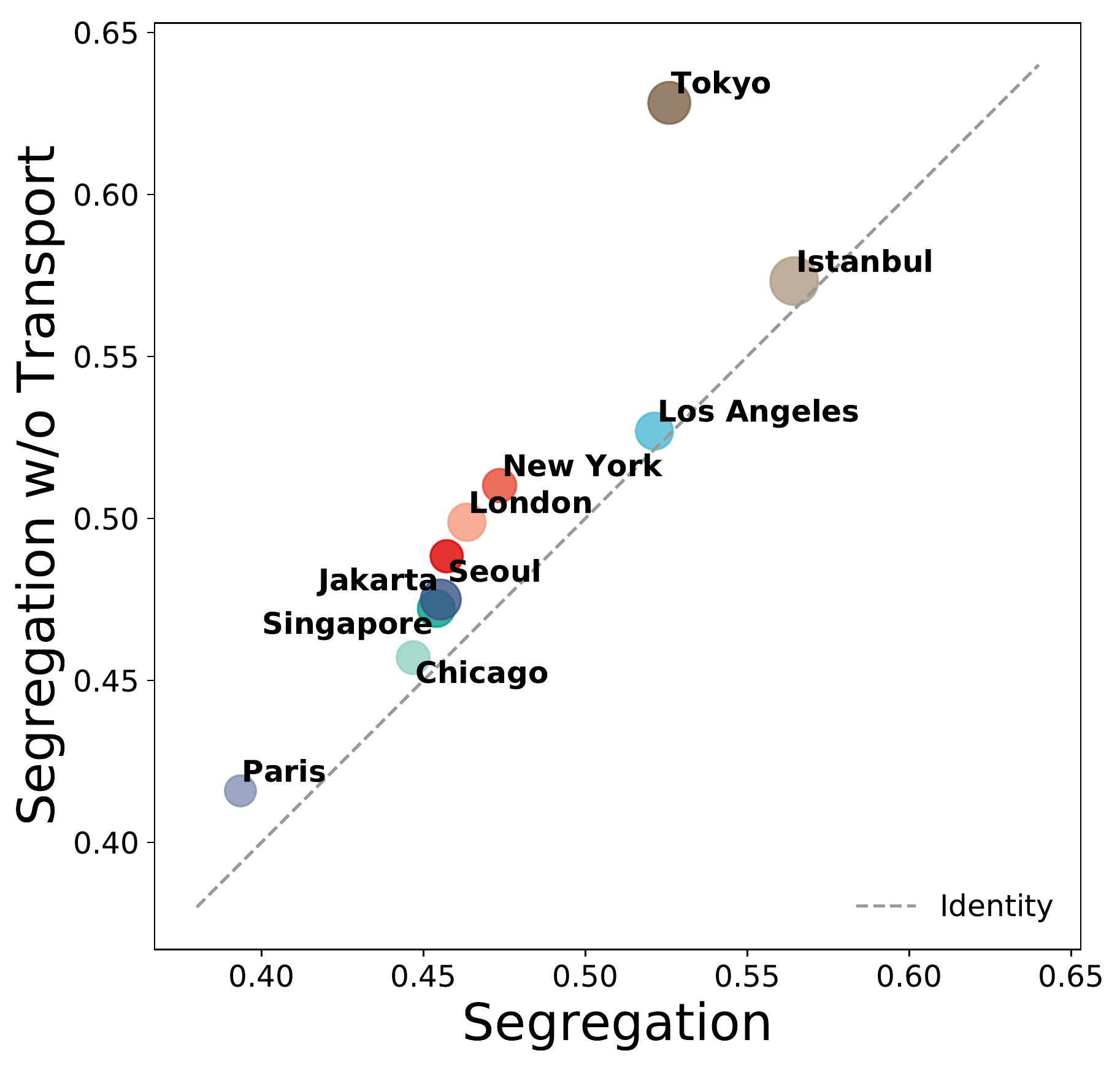} &
\raisebox{2.5cm}{(b)} \includegraphics[angle=0, width=0.45\textwidth]{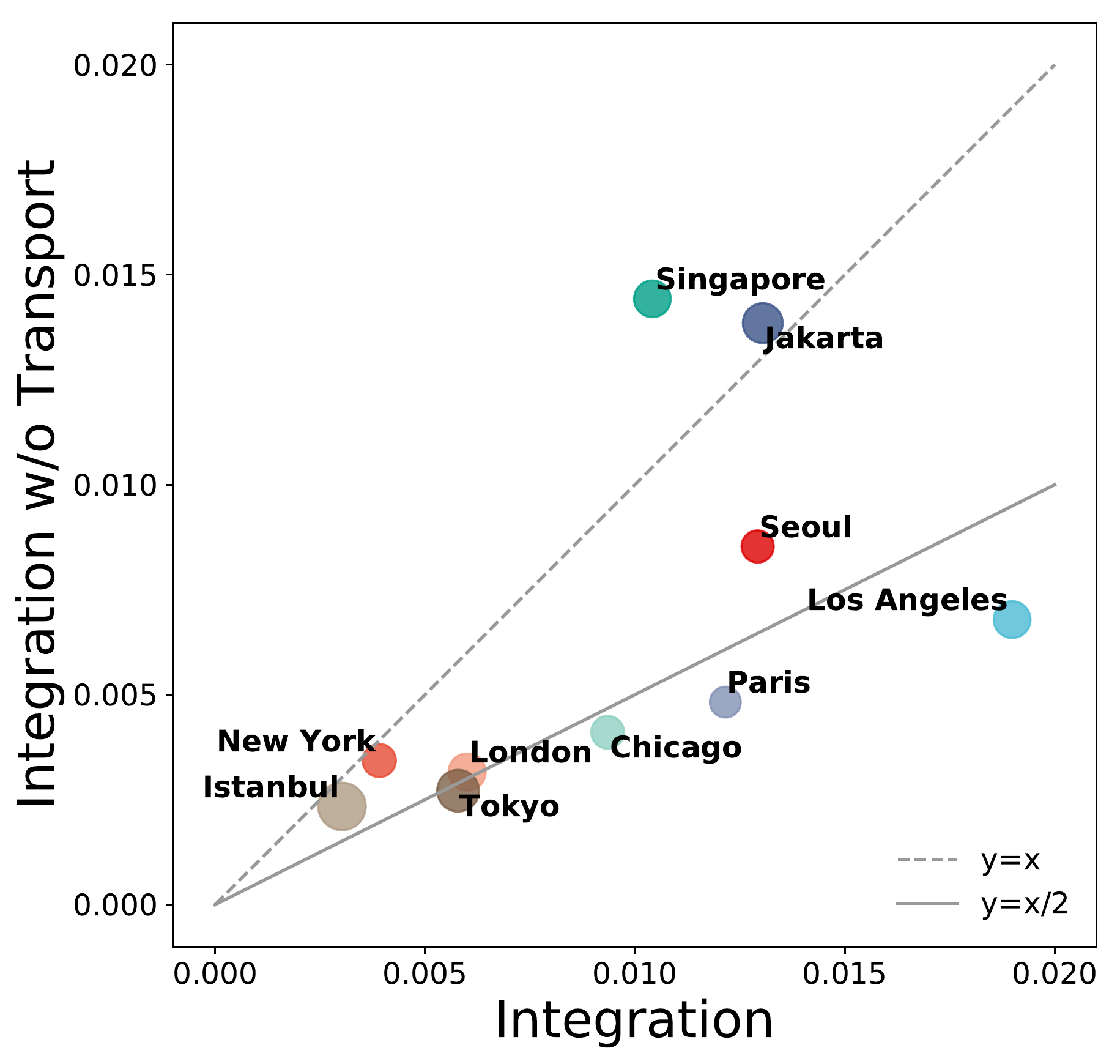} \\
\raisebox{2.5cm}{(c)} \includegraphics[angle=0, width=0.45\textwidth]{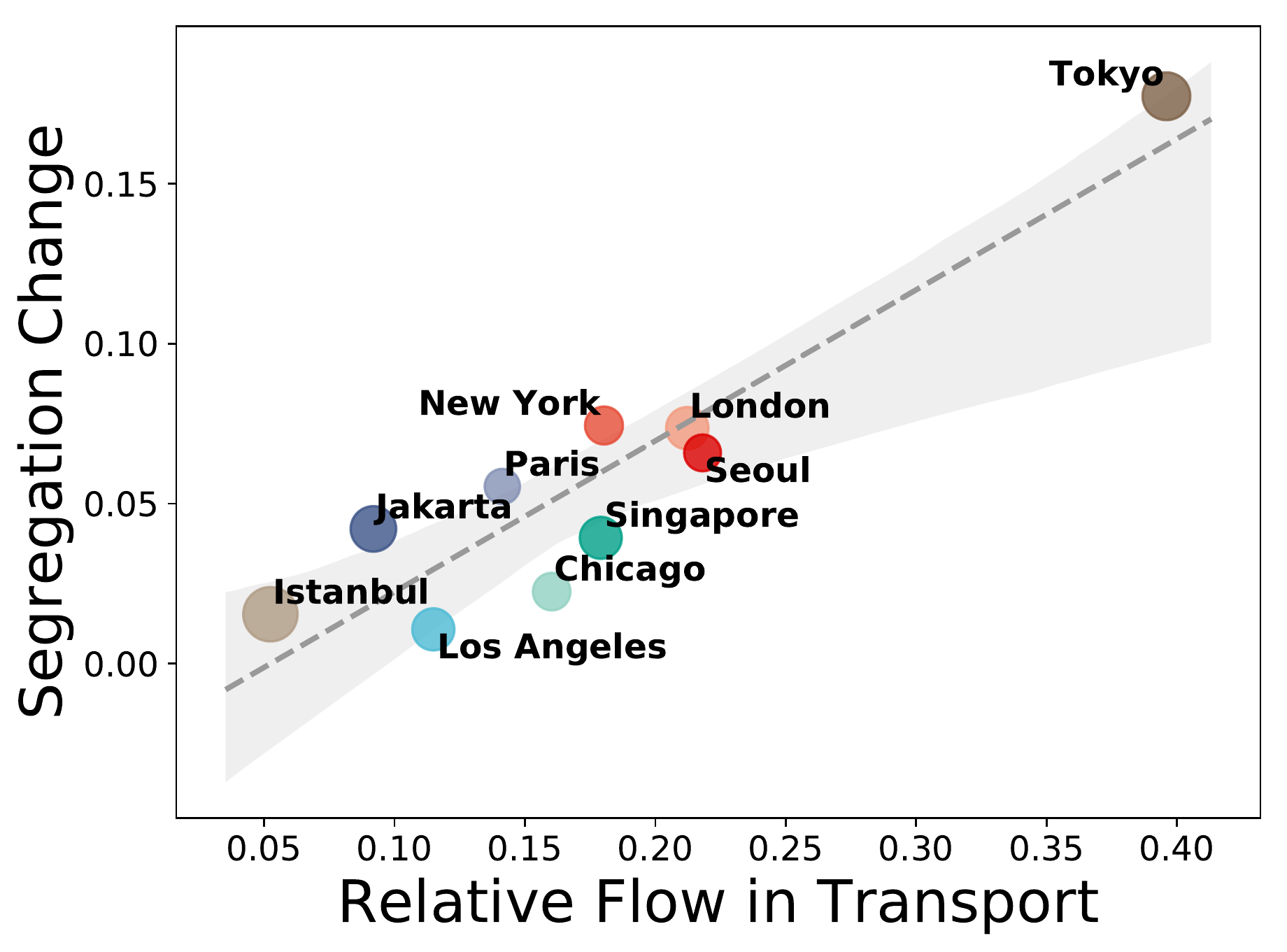} &
\raisebox{2.5cm}{(d)} \includegraphics[angle=0, width=0.45\textwidth]{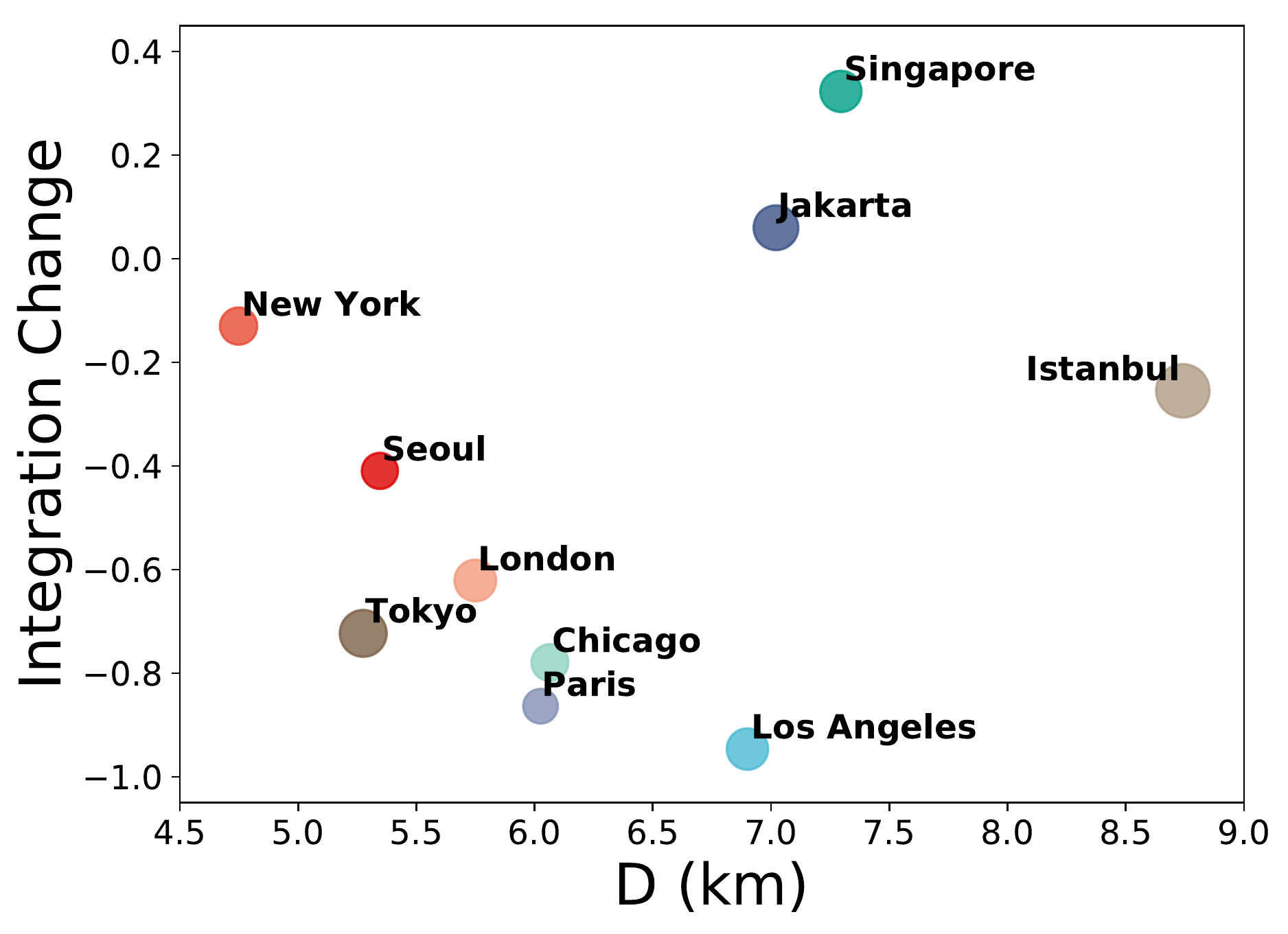} \\
\end{tabular}
\end{center}
\caption{
{\bf Illustrating the role of transport in building integration and reducing segregation.} As observed in Fig. \ref{fig5}b, the removal of the transport layer modifies significantly a city segregation and integration.
{\bf (a)} Segregation always increases after removing the transport layer.
{\bf (b)} Integration drops after removing the transport layer for some cities (that may reach values as smaller as the half of the initial value) but remains similar or even raises for other.
{\bf (c)} The raise in segregation grows linearly with the fraction of total flow represented by in the transport layer.
{\bf (d)} The relative change in integration ($GCE_{removed} - GCE_{full})/ GCE_{full}$ is not simply linked to the length of the connections cut: while for seven cities it seems to follow a trend similar to that pointed out in Fig.\ref{fig5}c, fort three cities where the average connection length of the transport layer is very large strongly deviate from this trend.
}
\label{fig_6}
\end{figure*}

It is natural observing how the transport layer represents the backbone of a city organization, but for some cities this effect is stronger than in others. To understand these differences, in Fig.~\ref{fig_6} we explore with more depth the difference in segregation and integration consequent to the removal of the transport layer. The effect is clear for the change in segregation (panels (a) and (c)): the increase in segregation. consequent to layer removal is proportional to how much flow pass though that layer. Things are, again, more complicated when we observe integration: for some city. the integration drops of $\approx 50\%$ without the transport layer, while for others (notably Singapore, Jakarta and Istanbul) integration is unchained, or even slightly increased, by the layer removal (panel (b)). These three cities have also the transport layer characterized by the longest average link distance (panel (d)), and while for the other seven cities one might have dared to see a trend, similar to that of Fig.~\ref{fig5} (c), linking higher drop in integration to longer connections, the presence of these three outliers suggests, another time, that microscopic details in the distribution of flows of a functional network can play a major role in determining its robustness and more general its organization.

\section*{Discussion}


Understanding how cities process information, here encoded by human flows, is of paramount importance for designing more efficient and smart urban systems and communities. By characterizing at the same the structural and the functional organization of 10 large-scale urban systems in terms of well defined and normalized measures of network integration and segregation, we have shown how network-based analysis can support, and further expand, ongoing discussions about and novel understanding provided by the ICT-data driven quantitative urbanism~\cite{louail2014mobile}. 

For growing cities, it is expected a transition from a monocentric to a polycentric organization, characterized by a sub-linear growth of the number of hotspots with population~\cite{louf2013modeling}. Similarly, for both urban structural and functional networks, we provide evidence that large polycentric cities which are characterized by a larger number of hotspots(although being the growth sub-linear they have a smaller fraction of hotspots as shown in Supplementary Fig.~\ref{SI:city_size} d), appear to be more segregated and less integrated than smaller, and monocentric, cities. We have highlighted, however, that a city can be much more integrated than what expected by its size if it display a low flow-hierarchy~\cite{bassolas2019hierarchical} and thus has more direct connections between central and marginal areas.

From a modeling perspective, we discover that many features of complex megacities can be understood from simple mechanisms related to geometric constraints and city's characteristic size, with larger cities tending to be more segregated and less integrated. Small-world models based on lattices with with long-range connections seems to be a good candidate to reproduce the many salient features measured from empirical data. However, the interplay between heterogeneities in the distribution of flows and spatial constraints might be responsible for the emergence of peculiar integrated/segregated structures that might be reflected in the functional organization of the city. Future research in this direction, including a wider spectrum of urban and non urban systems, is required to gain more insights on this matter.

Lastly, from a more methodological perspective, our analysis highlights the importance of data sources for the analysis of the interplay between the city and its main users, i.e., the citizens. Thanks to the unique dataset of anonymized movements provided by Foursquare and the easy access to street data~\cite{boeing2017osmnx}, we have been able to gain novel insights on urban and human behavior in terms of interaction between structure and functional organization of the system. The availability of activity-aware information, in particular, allowed the analysis of attacks targeted towards specific types of activities which unraveled the fundamental importance of transport as integrator a urban system. This result is specially relevant for policy and decision-making in time of crisis, provide new quantitative tools that allow one to identify a limited set of activities (commercial, restaurants, leisure) which can be prioritized or temporary limited to achieve a desired amount of human flows integrated across the city.

\section*{Methods}
\label{sec:methods}

\heading{Geographic coarse-graining} We reconstruct the flows network by aggregating data over areal units of 500m $\times$ 500m, in all 10 cities considered. Flows are reconstructed from subsequent anonymized check-ins into Foursquare venues, ignoring the order (undirected network). Flows inside the same area have been integrated into a self-loop link only if the check-ins were between two different locations. Subsequent check-ins in the same location have been excluded from the analysis. 

\heading{Temporal stratification} We decouple the functional use of a city i) at different hours of the day (morning, midday, afternoon, night, overnight), and ii) in different months of the year.

\heading{Activity stratification} We use Foursquare's rich system of categories and manually associate them to a reduced number of macro-categories (food, lodging, tourism, work, religion, services, education, health, sport, transport, entertainment, leisure, public, housing and commercial). We do not use \cite{FoursquareGetVenue}, except for venue icons in Fig.~\ref{fig:first_figure}. The few categories that did not fit any macro-category have been labelled as `unknown'. These categories allow us to build ``activity-aware multilayer networks'', where activities of different types are associated to different layers of our model. Flows between activities of the same macro-category are encoded by intra-layer links, while flows between different categories are encoded by inter-layer links.

\heading{Measuring functional integration} We measure to which extent a network is integrated in terms of communication, i.e., how efficient nodes are, on average, in exchanging information.
Given two areal units $i$ and $j$ we can reasonably assume that the efficiency $\epsilon_{ij}$ in their communication is inversely proportional to their distance $d_{ij}$. 
If $d_{ij}$ is a topological distance, counting the number of links in a shortest-path from $i$ to $j$, our assumption means that the longer the path a piece of information has to travel, the more inefficient will be the communication, since the probability that the message is corrupted along the way increases.
A global descriptor of the topological communication efficiency~\cite{latora2001} of a city is then the average pairwise efficiency of its nodes
\begin{equation}\label{eq:topo-eff}
E = \frac{1}{N(N-1)} \sum_{i\neq j} \frac{1}{d_{ij}}.
\end{equation}

Many real systems, among which those analyzed in this paper, are better described by weighted networks which encode the additional information on the strength of connections. 
Distances are very different in these networks: if the flow between two nodes is large, their distance should be, intuitively, small.
Weighted shortest-path distances minimize the sum of \emph{costs} along all paths between pairs of nodes.
In a flow network with edge weights representing the intensity of the connections, the costs of edges are the inverse of weights.
Unfortunately, \eqref{eq:topo-eff} cannot be effortlessly generalized to weighted networks, since it depends on the scale of weights.
Although Latora and Marchiori proposed a weighted efficiency descriptor in~\cite{latora2003economic}, it has been shown~\cite{bertagnolli2020quantifying} that finding the ideal proxy $G_{\text{ideal}}$ of a network $G$ for the normalization of the weighted $E(G)$ is often ambiguous.
Here we use the normalized GCE proposed in~\cite{bertagnolli2020quantifying}, where $G_{\text{ideal}}$ is the result of a physically-grounded procedure, which is independent from the scale of flows and from any metadata or the lack thereof.
The idea is that each (weighted) shortest-path in the network has a length, which is the sum of links costs along the path and a total flow, which is the sum of the links weights. 
These path flows $\phi_{ij}$ are strictly positive for each pair of nodes $(i, j)$ in a connected network and can be added to the original network as an artificial flow. In other words, the artificial links represent shortcuts between pair of nodes -- they deliver the total flux through a shortest-path from origin to destination in one topological-step -- and the network $G_{\text{ideal}}$ resulting from this enrichment procedure enables a correct normalization of $E$.

\heading{Measuring functional segregation} A usual measure of network segregation, quantifying how strongly the units are organized in into $M$ non-overlapping blocks, is the modularity~\cite{Newman2004b}
\begin{equation}
    Q = \sum_{u \in M} \left[e_{uu} - \left(\sum_{v \in M} e_{uv}\right)^{2}\right]
\end{equation}
where $e_{uu}$ is the proportion of links inside module $u$, while $e_{uv}$ accounts for the connectivity between two distinct modules $u$ and $v$.
More specifically, our measure of segregation is the maximum value of the modularity that we find using the Louvain algorithm~\cite{blondel2008fast}. We also verify that the observed modularity is significant, by comparison with the values of $Q$ computed over an ensemble of configuration models obtained reshuffling the  network.
Finally, note that here, instead, we used the weights defined by flows. Values of $Q$ for weighted and unweighted networks are indeed comparable, as opposite to what discussed above for $E$, and using weights here allowed us to better discern the characteristics of different layers.  

\heading{Synthetic network models} We use two standard spatial network models for our analysis. 

We first consider a class of networks characterized by small average geodesic distance: the Watts-Strogatz (WS) model. Starting from a regular graph, e.g., a two-dimensional lattice, each link has a probability $p$ of being \textit{rewired}, that is removed and re-placed randomly in the network. If $p$ is large the resulting WS network will look more like an ER random graph than the original lattice. WS networks are also highly clustered, where nodes tend to form closed triangles. WS model are usually referred to as \textit{small-world} networks.

Alternatively to WS, we study also the simplest network model actively involving the spatial dimension model is the random geometric network (RGN), where nodes randomly distributed in space are connected if they are closer than a fixed threshold distance. The RGNs share many important properties with regular lattices, in particular they are not ``small world''. For this reason, similarly to the WS case, here also for the RGN we perform a rewiring with probability $\alpha$.

\heading{Acknowledgements} The authors thank Foursquare for granting access to the data set used in this study and acknowledge Matthew Kamen, Renaud Lambiotte, Jesse Lane, Anastasios Noulas, Cecilia Mascolo, Vsevolod Salnikov, Sarah Spagnolo and Adam Walksman for organizing the Future Cities Challenge. The authors acknowledge Giuseppe Lupo and Valeria d'Andrea for fruitful discussions.

\section*{Author contributions}
RG and MDD designed research. RG, GB and MDD performed the research and wrote the paper.

\section*{Competing interests}
The authors declare no competing interests.

\section*{Data availability statement}
Aggregated data allowing to reproduce this papers' results are available from the authors upon request.



\clearpage

\bibliographystyle{unsrtnat} 

\begin{small}
\bibliography{biblio}
\end{small}

\clearpage

\clearpage
\section*{Supplementary Materials}

\setcounter{figure}{0} 


\begin{table}[h!t]
\renewcommand{\tablename}{Supplementary Table}

\begin{tabular}{l|r|r|r}
City & \#Venues & \#check-ins & L (km) \\
\hline
Chicago & 13904 & 10629110 & 18.9 \\
Istanbul & 113752 & 13083383 & 39.3 \\
Jakarta & 21813 & 9281181 & 27.6 \\
London & 22689 & 10146880 & 24.0 \\
Los Angeles &  15868 & 10362146 & 23.6\\
New York City & 32971 & 11048584 & 19.3\\
Paris & 13588 & 9521723 & 16.6 \\
Seoul & 15545 & 9347489 & 18.1\\
Singapore & 23324 & 9691517 & 23.5\\
Tokyo & 57810 & 11545155 & 30.4\\
\end{tabular}
\caption{\label{fig:table_data}
\textbf{Foursquare data set extensive characteristics.} The figures here are aggregated for all layers, hours, and comprise all 24 months. The linear size $L$ is here estimated as the square root of the total area covered by the data after the aggregation into squares of $500m \times 500m$.}
\end{table}


\begin{figure*}[!t]
\renewcommand{\figurename}{Supplementary Figure}
\begin{center}
\begin{tabular}{cc}
\raisebox{2.5cm}{(a)} \includegraphics[angle=0, width=0.45\textwidth]{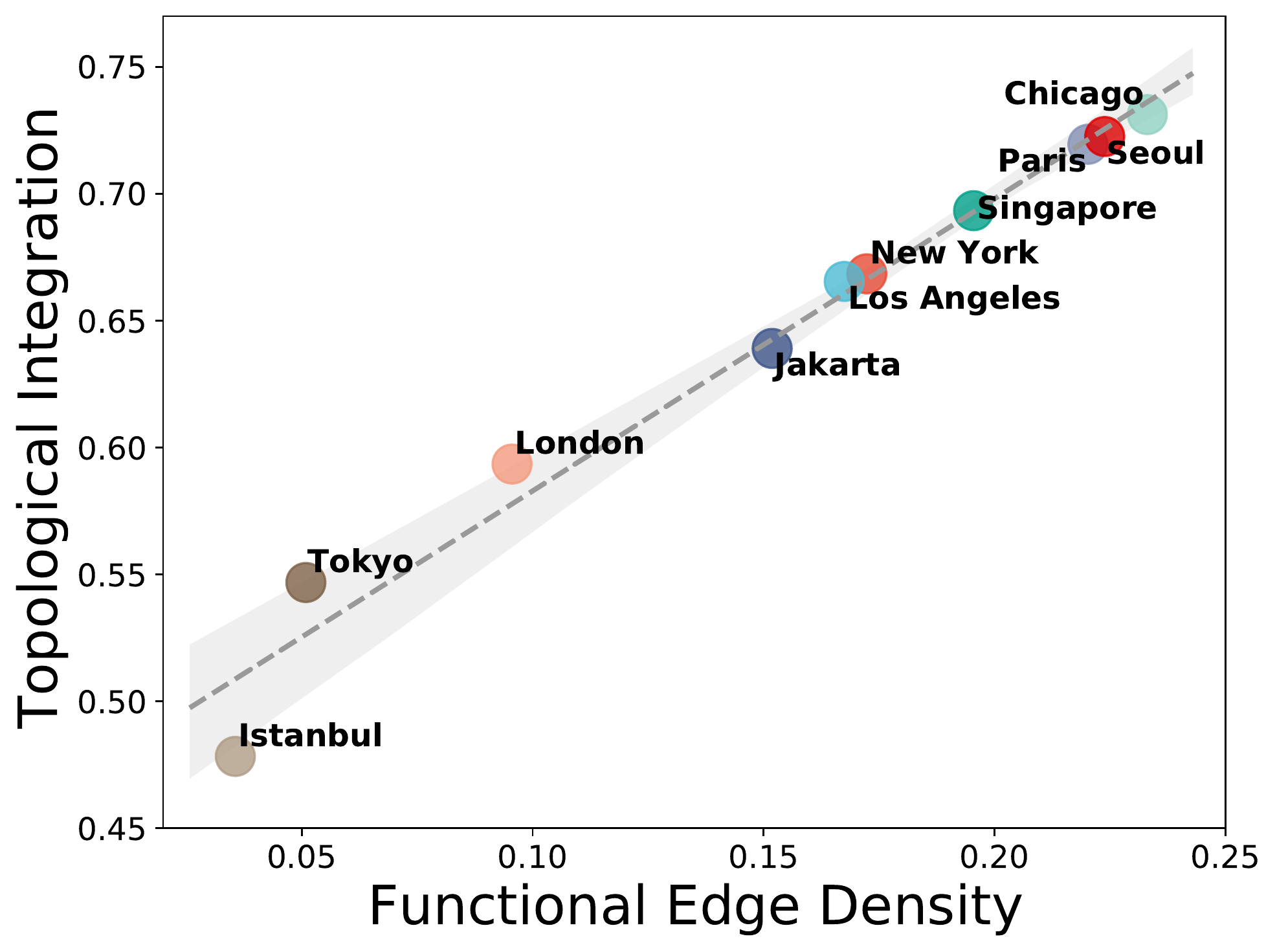} &
\raisebox{2.5cm}{(b)} \includegraphics[angle=0, width=0.45\textwidth]{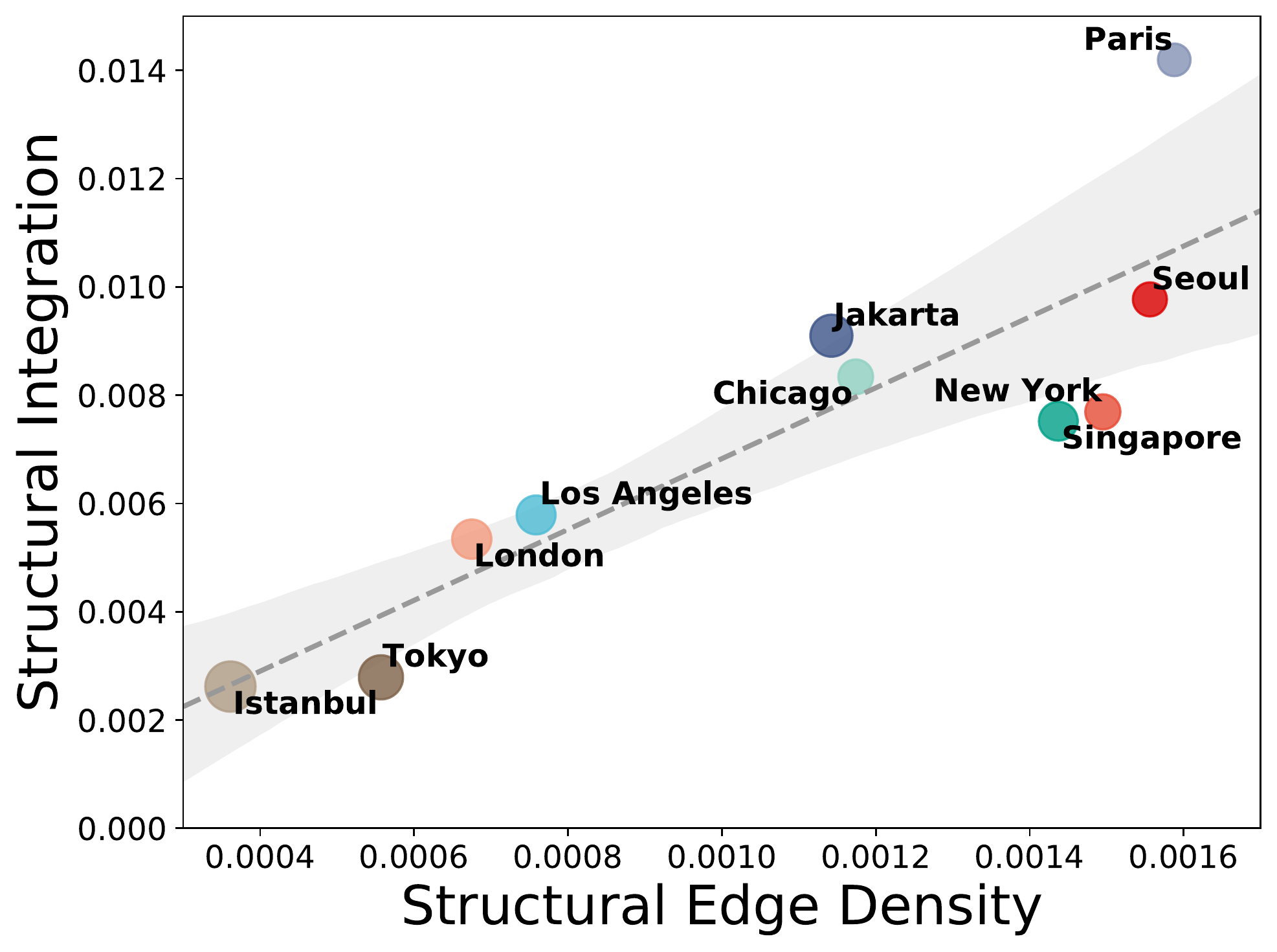}
\end{tabular}
\end{center}
\caption{
{\bf Dependance of integration of topological networks on edge density} 
{\bf (a)} The value of integration for the topological functional networks is strictly linked to the edge density. As expected for topological networks, the larger the edge density the larger the integration.
{\bf (b)} The integration for the structural networks is also strongly related to edge density. 
\label{SI:GCE_ed}
}
\end{figure*}

\begin{figure*}
\renewcommand{\figurename}{Supplementary Figure}
\begin{center}
\begin{tabular}{cc}
\raisebox{2.5cm}{(a)} \includegraphics[angle=0, width=0.45\textwidth]{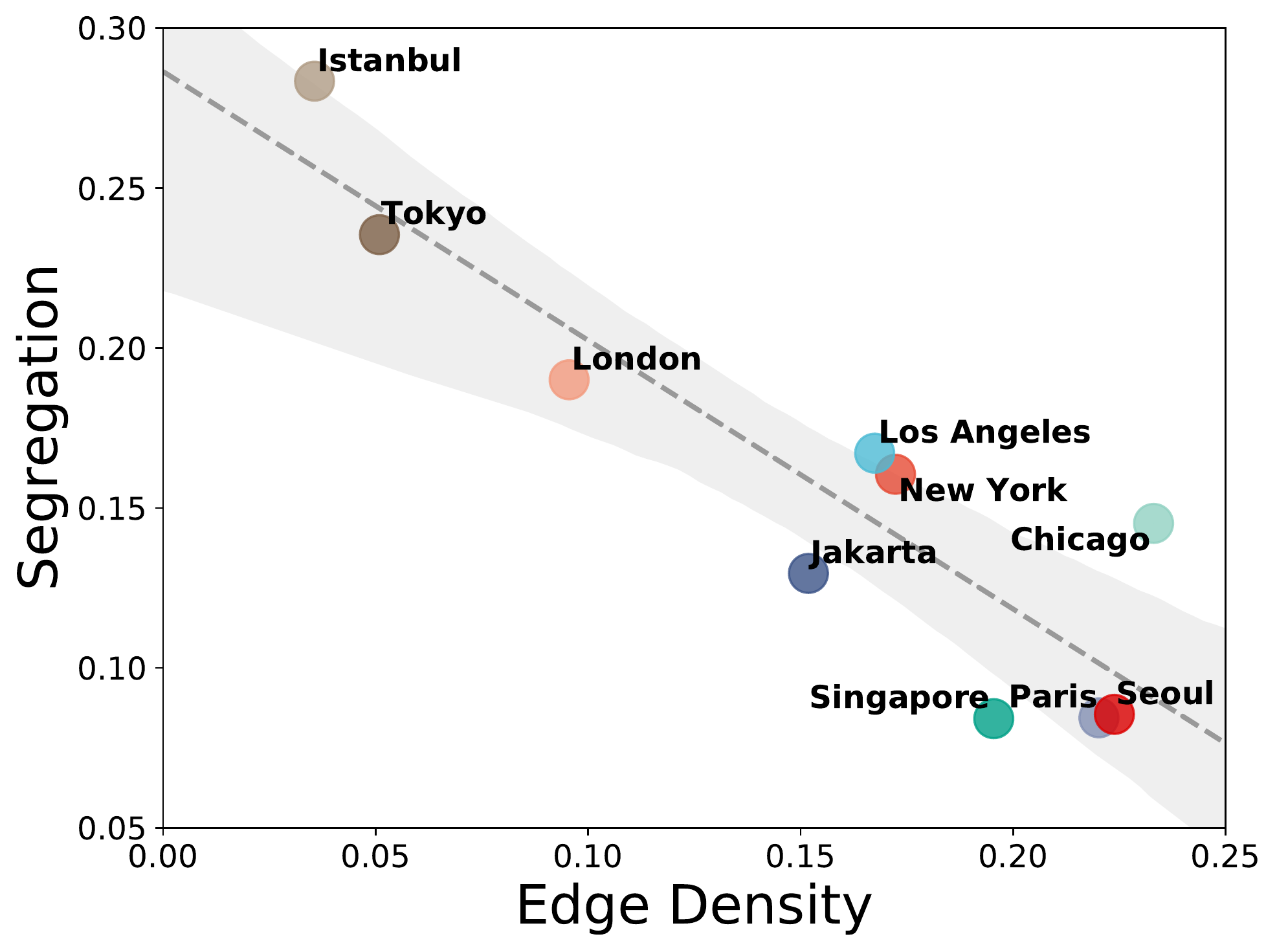}&
\raisebox{2.5cm}{(b)} \includegraphics[angle=0, width=0.45\textwidth]{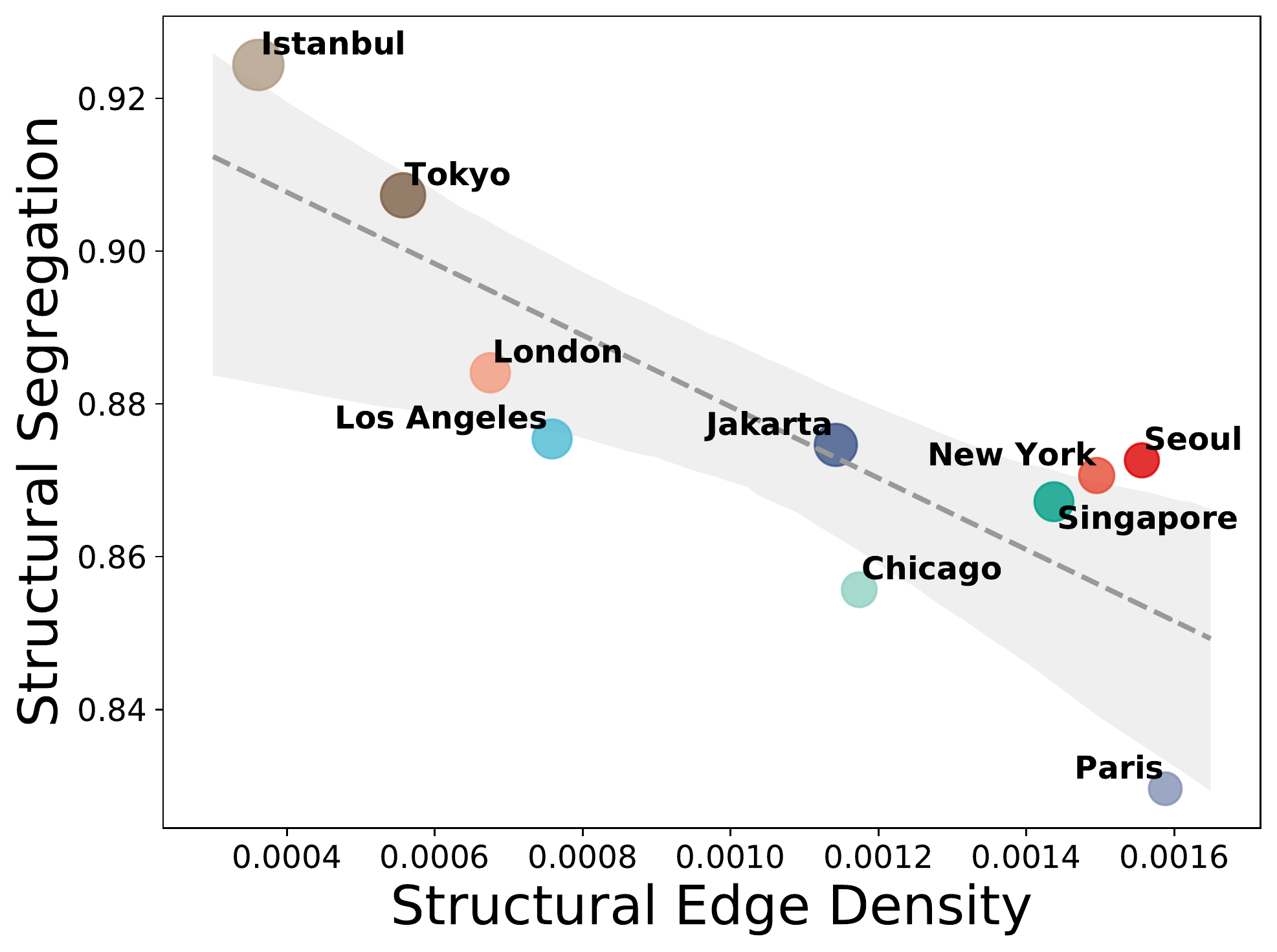} \\
\end{tabular}
\raisebox{2.5cm}{(c)} \includegraphics[angle=0, width=0.45\textwidth]{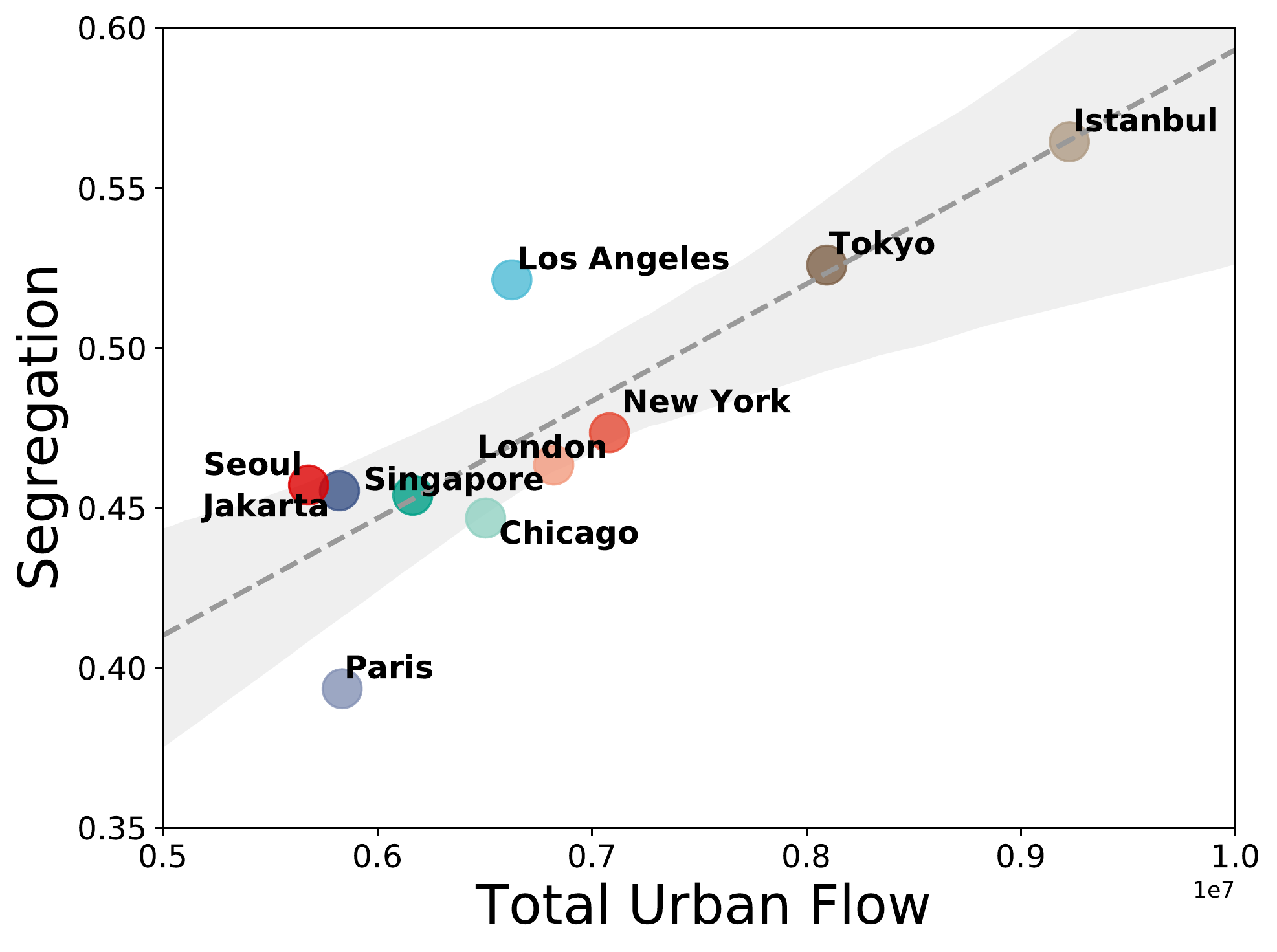} 
\end{center}
\caption{
{\bf Understanding Structural and Functional Segregation.} Similarly to what illustrated in Fig. 4, here we link the values of  segregation for the structural network derived from Open Street map and the functional (topological and. weighted) networks derived from the Foursquare flows. In this case, the values observed for the three networks are not consistent one the other, suggesting that an improved and correctly normalized definition of segregation is still needed.
{\bf (a)} The value of segregation for the topological functional networks  is anti-correlated to the edge density, but less tightly than what observed for Integration.
{\bf (b)} Similarly, segregation for the structural network grows as edge density increases.
{\bf (c)} The value of segregation for the weighted functional network seems instead to be linked to the total flow recorded in the city, i.e. is the sum of all weights in the network.
}
\label{SI:segregation}
\end{figure*}

\begin{figure*}
\renewcommand{\figurename}{Supplementary Figure}
\begin{center}
\begin{tabular}{cc}
\raisebox{2.5cm}{(a)} \includegraphics[angle=0, width=0.45\textwidth]{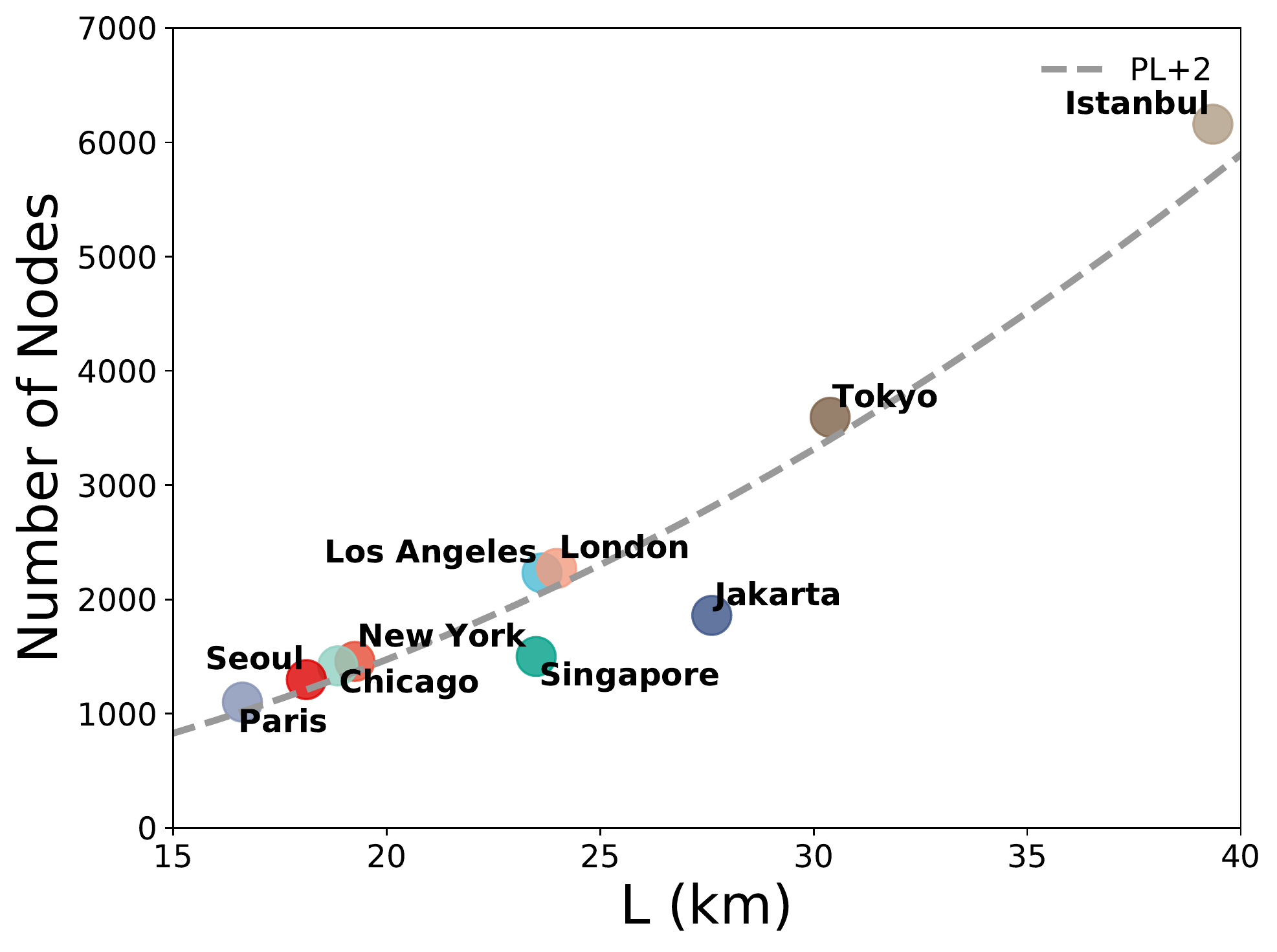}&
\raisebox{2.5cm}{(b)} \includegraphics[angle=0, width=0.45\textwidth]{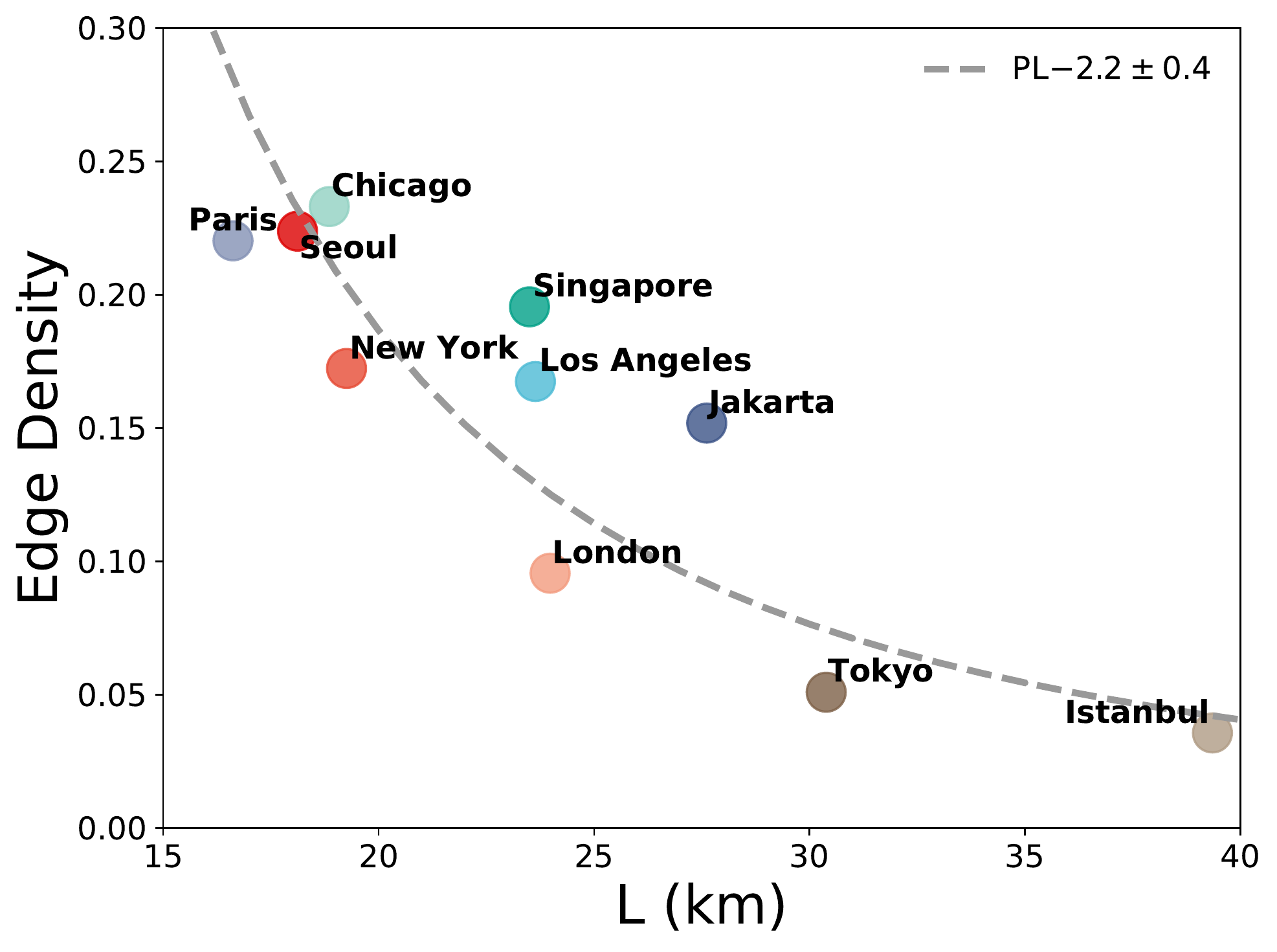} \\
\raisebox{2.5cm}{(c)} \includegraphics[angle=0, width=0.45\textwidth]{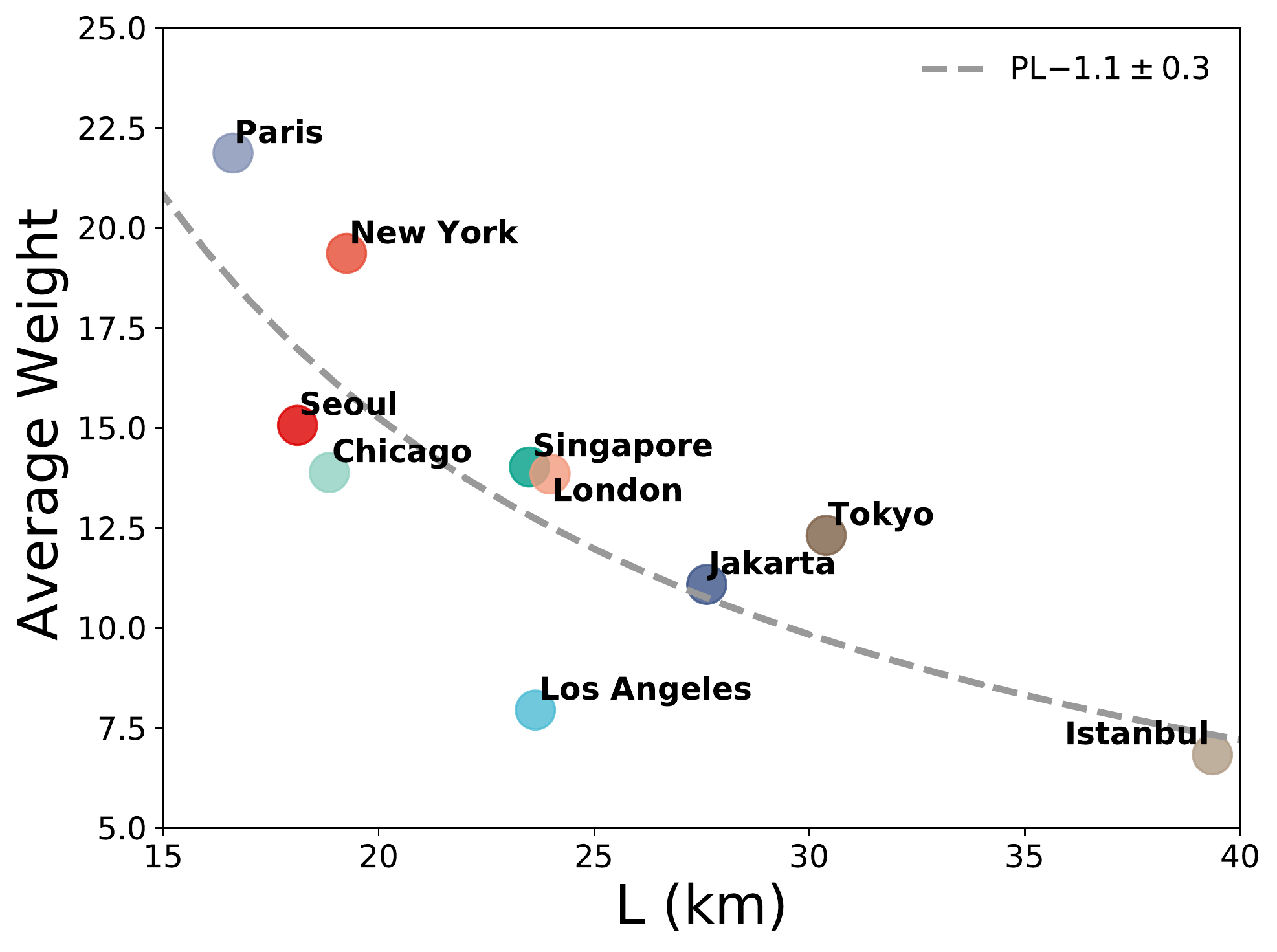} &
\raisebox{2.5cm}{(d)} \includegraphics[angle=0, width=0.45\textwidth]{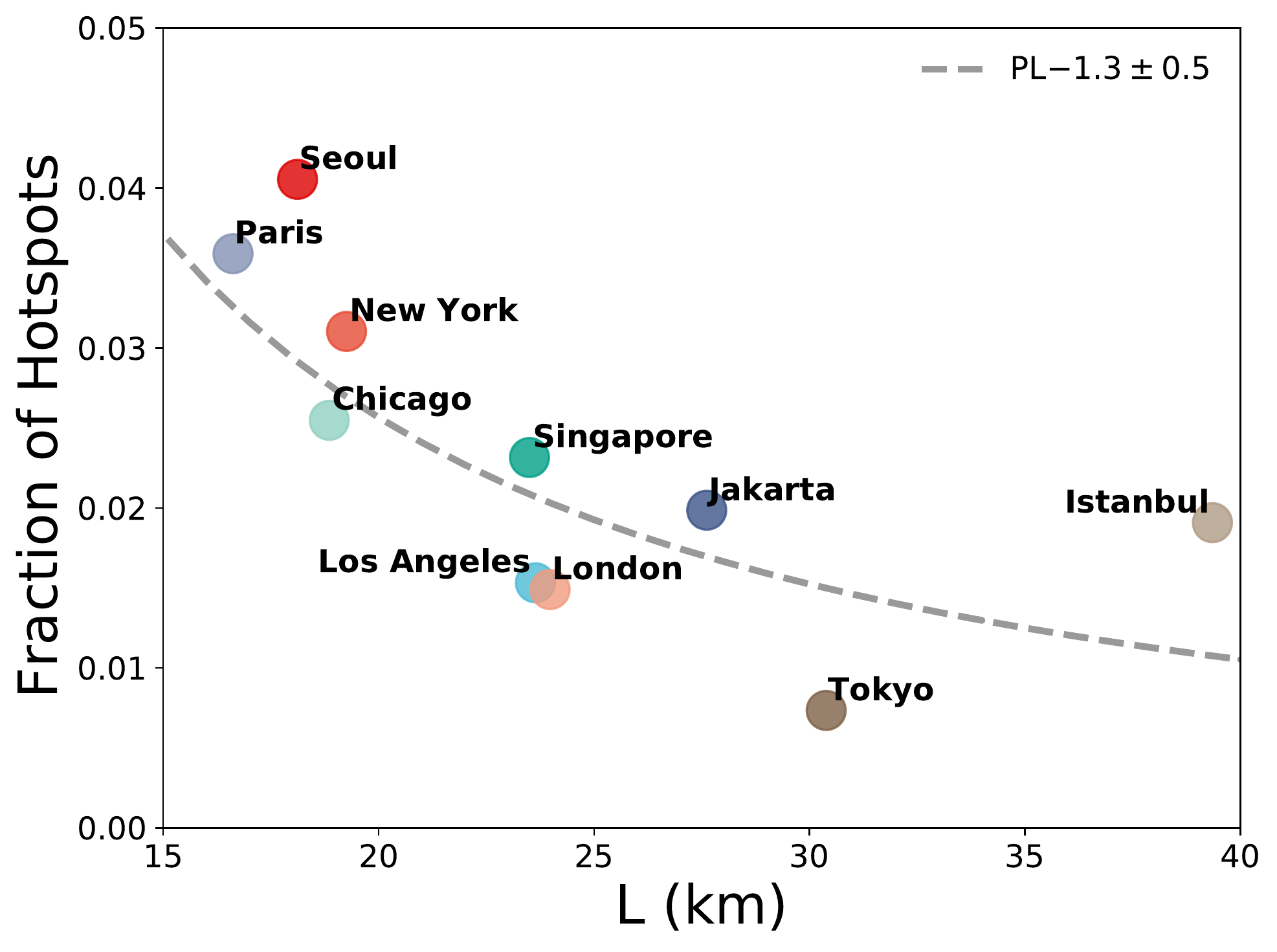} \\
\end{tabular}
\end{center}
\caption{
{\bf Connecting network properties with urban scaling.} Here we show how the network indicators we extracted from the Foursquare data depend upon city dimensions in terms of $L$, that is computed as the  square root of the surface area included in the data provided. As we observed in Supplementary Fig.\ref{SI:segregation}, functional segregation appears to be proportionate to the total weight of a city. The total weight can be decomposed to the product of three factors: $W_{tot}= N^2 \cdot ed \cdot \langle w\rangle$. In the first three panels we illustrate the scaling behavior for these three quantities (number of nodes, edge density and average weight respectively).
{\bf (a)} Since we have built the network by coarse graining on a regular grid, it is natural that number of nodes is naturally proportionate to the square of $L$, i.e. the surface area.
{\bf (b)} The edge density decreases for larger cities, which leads to higher values of topological segregation and integration as the city grows. 
{\bf (c)} Also the average weight of links decreases for larger cities, a factor contributing to a smaller values of segregation as the city grows.
{\bf (d)} A final insight on the scaling properties of cities can be extracted by observing that as the size $L$ of the city grow, the fraction of area that is represented by hotspots obtained with the LouBar method~\cite{louail2014mobile}  decreases.
All dashed lines represent the best fit for a power-law scaling. Given the limited number of points and decades the values have to be considered only as a rough indication which we include in this figure as we hope might inform further studies.
}
\label{SI:city_size}
\end{figure*}

\begin{figure*}
\renewcommand{\figurename}{Supplementary Figure}
\begin{center}
\begin{tabular}{c}
\includegraphics[angle=0, width=0.75\textwidth]{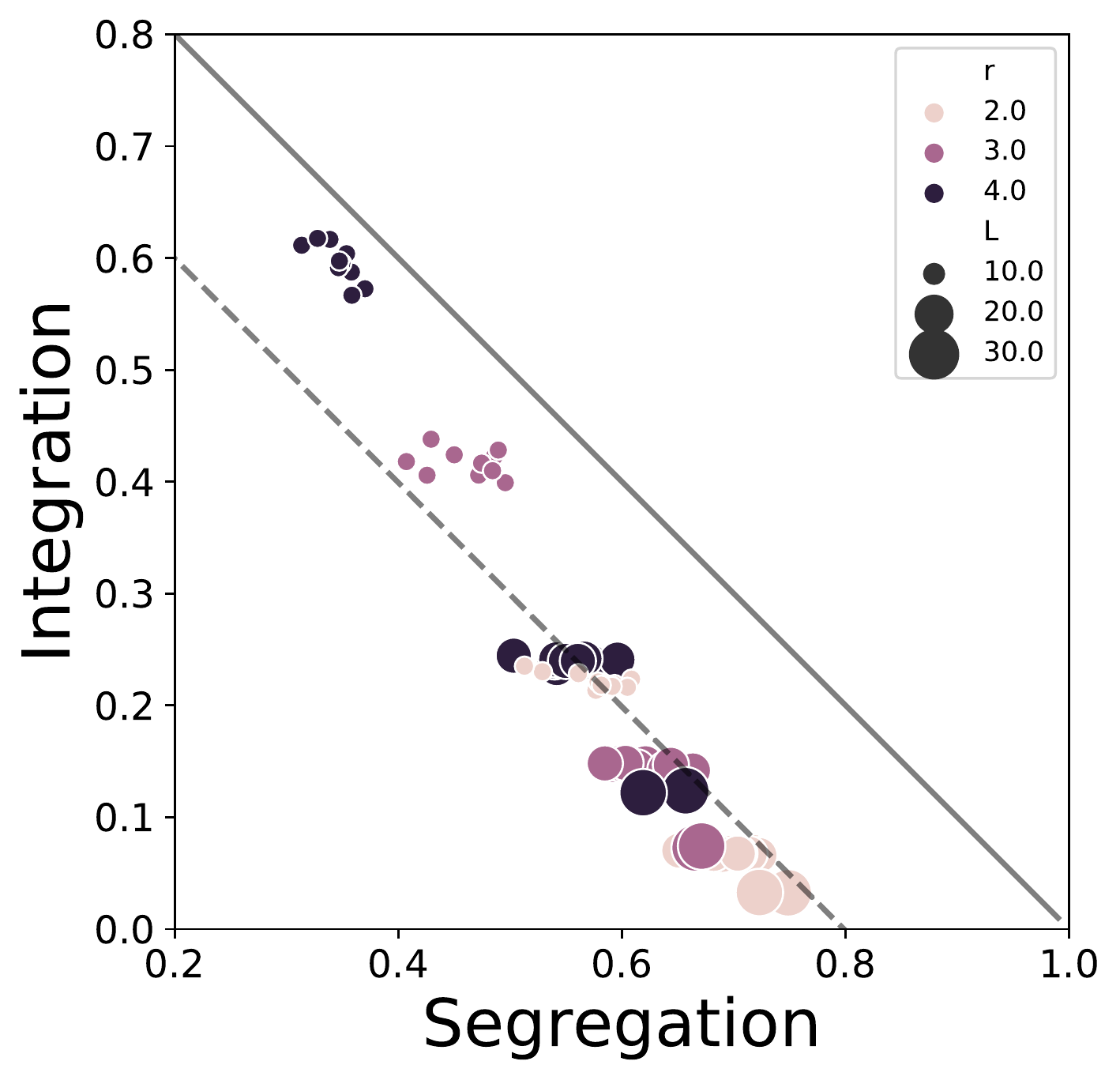}
\end{tabular}
\end{center}
\caption{
\textbf{Segregation and Integration of Random Geometric Networks of different sizes.} In this paper, we generate RGNs by i) throwing $N$ nodes in random locations in a square of edge $L$; ii) connecting all nodes i,j with distance $d(i,j) < r$; iii) rewiring a fraction $\alpha$ of edges. Here, to study the effect of size, we generate networks with identical node density $N/L^2$ and with no rewiring $\alpha = 0$. For each value of  $L$ and $r$ we averaged the values of segregation (modularity $Q$ and integration $GCE$). The result show that, in this scenario, segregation and integration are strongly anti-correlated. High integration is attained for small networks ($L = 10$) with large $r$, while the opposite yields high segregation.
}
\label{SI:RGG_scaled_QvsGCE}
\end{figure*}

\begin{figure*}
\renewcommand{\figurename}{Supplementary Figure}
\begin{center}
\begin{tabular}{c}
\raisebox{2.5cm}{(a)} \includegraphics[angle=0, width=0.8\textwidth]{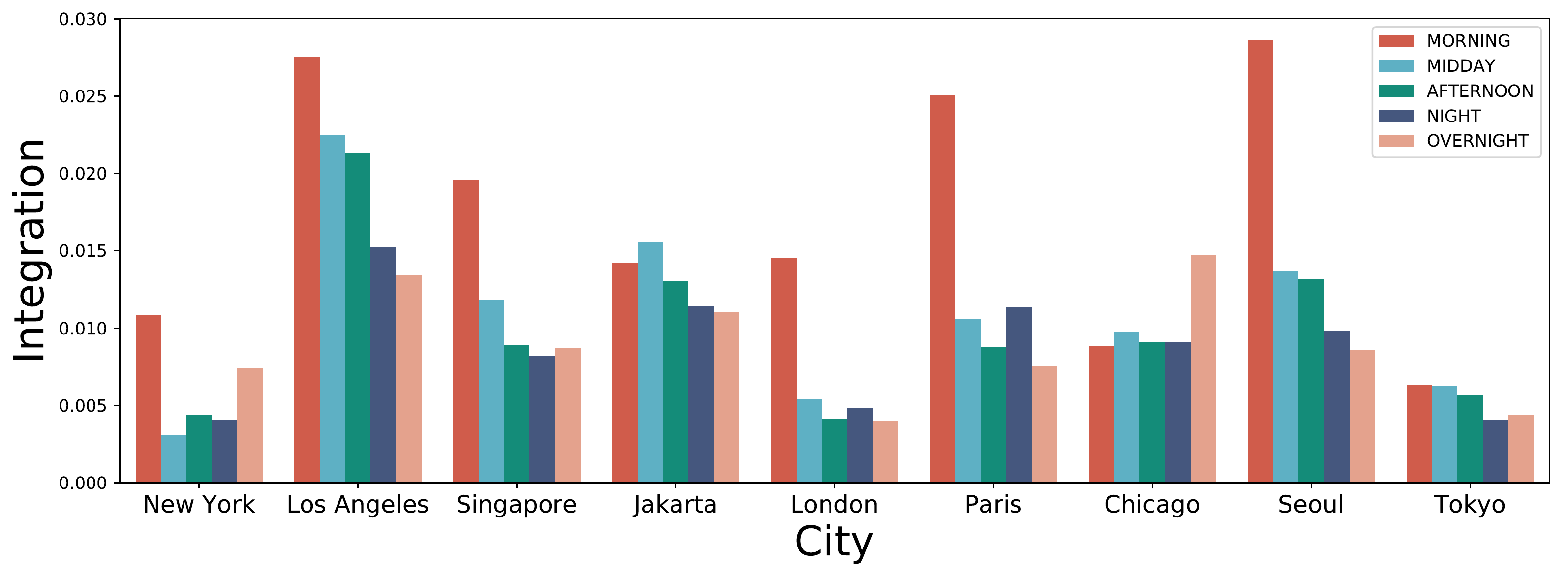}\\
\raisebox{2.5cm}{(b)} \includegraphics[angle=0, width=0.8\textwidth]{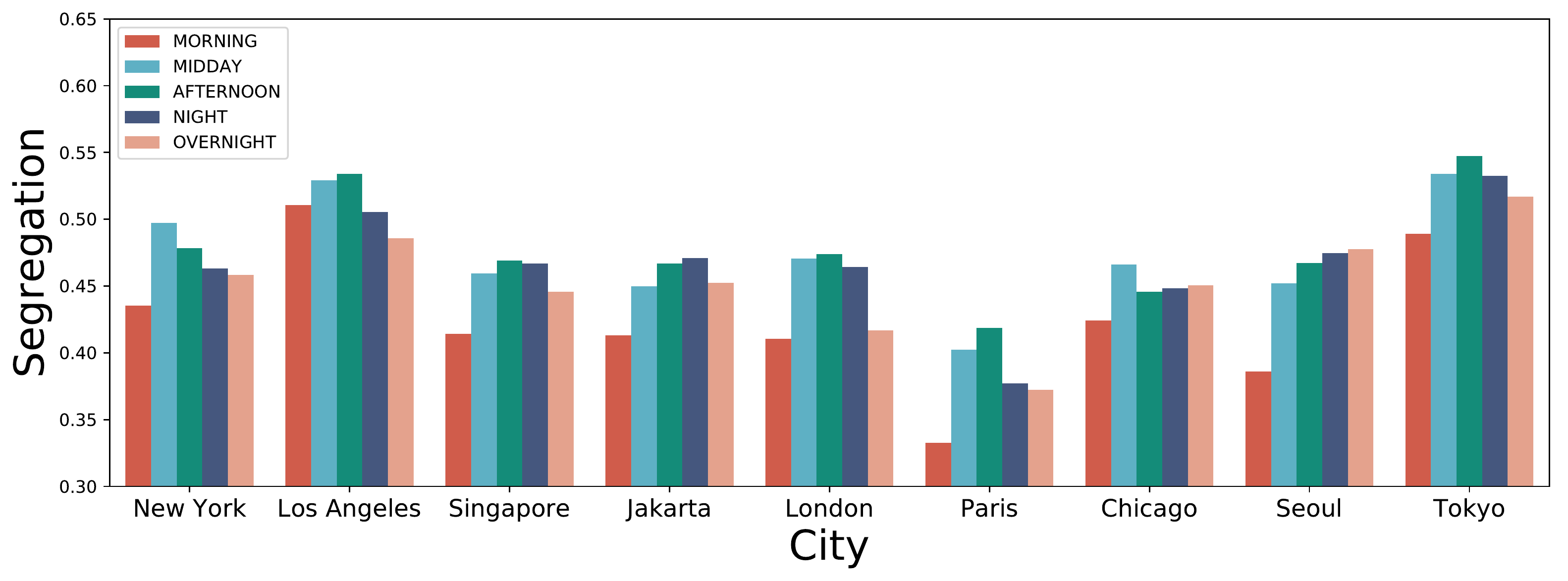} 
\end{tabular}
\end{center}
\caption{\bf
Segregation and Integration at different hours of the day.
}
\label{SI:hours}
\end{figure*}

\begin{figure*}
\renewcommand{\figurename}{Supplementary Figure}
\begin{center}
\begin{tabular}{cc}
\raisebox{2.5cm}{(a)} \includegraphics[angle=0, width=0.45\textwidth]{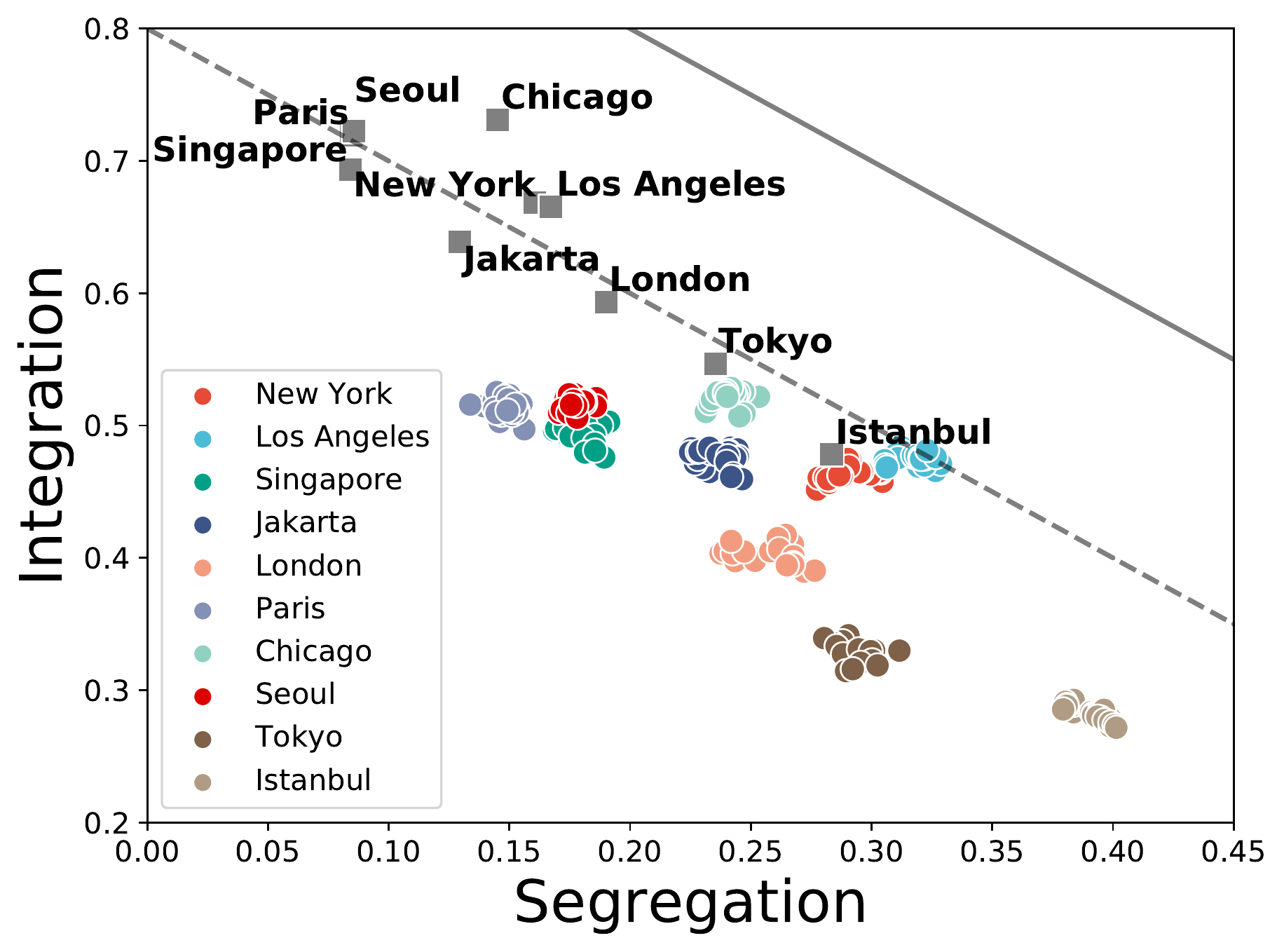}&
\raisebox{2.5cm}{(b)} \includegraphics[angle=0, width=0.45\textwidth]{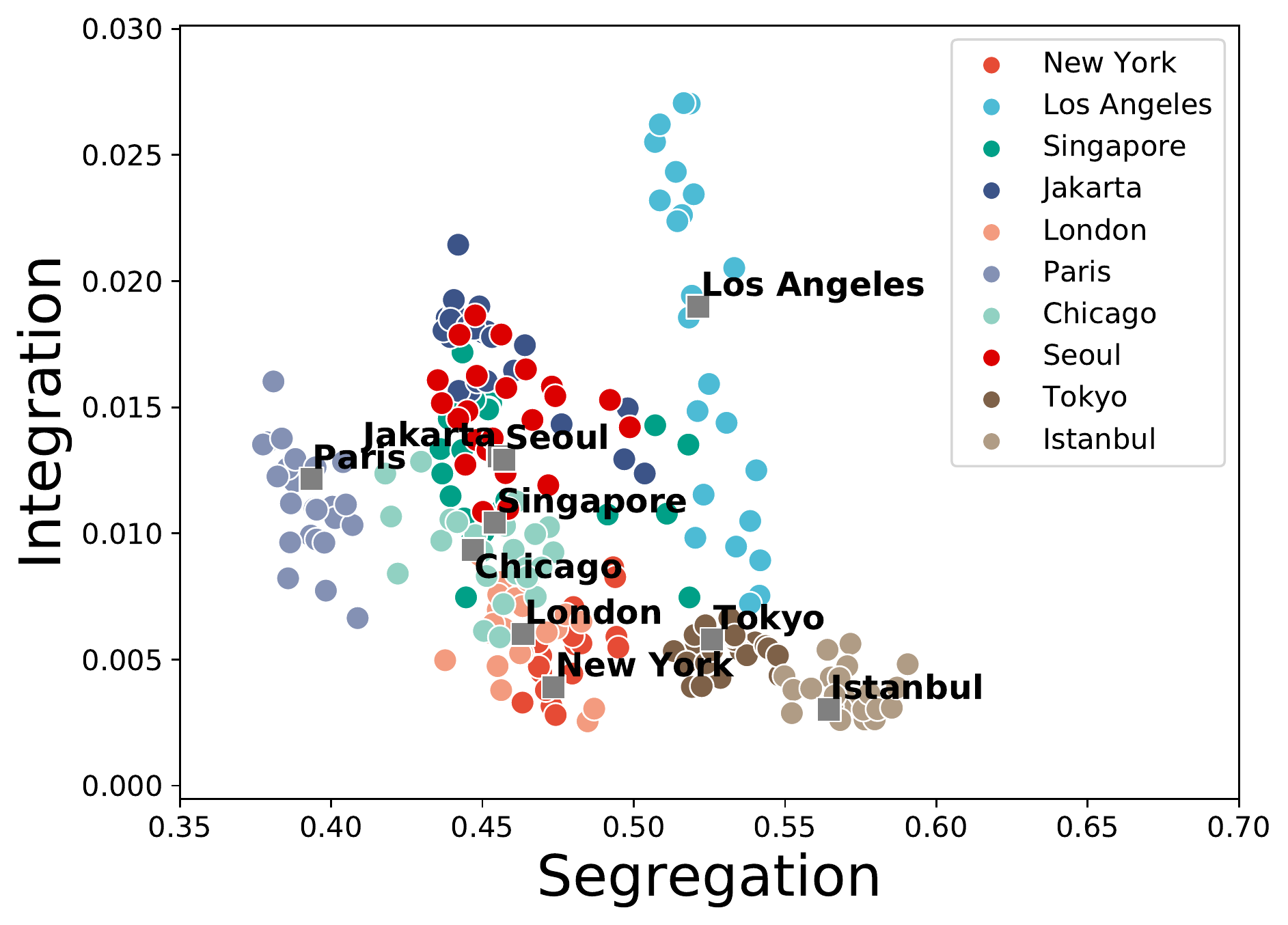} \\
\end{tabular}
\end{center}
\caption{
{\bf Integration and segregation for topological network and disaggregated by month.}  Flows are stratified according to different months (multiple points), while the grey square letter of a city name falls in correspondence of the values for the whole dataset.
{\bf (a)} Topological network. Remarkably, the values for topological network extracted by single months exhibit a large deviation from.  time, ranging from a functional organization resembling random geometric networks, suggesting that monthly data would be too undersampled for making an analysis based only on the topological features of the networks.
{\bf (b)} Weighted network. In this case, the more rich information captured by nodes allow to compare values for a single month (colored dots) to those aggregated over the whole period considered (grey square).
}
\label{SI:months}
\end{figure*}


\begin{figure*}
\renewcommand{\figurename}{Supplementary Figure}
\centering
\includegraphics[width=\textwidth]{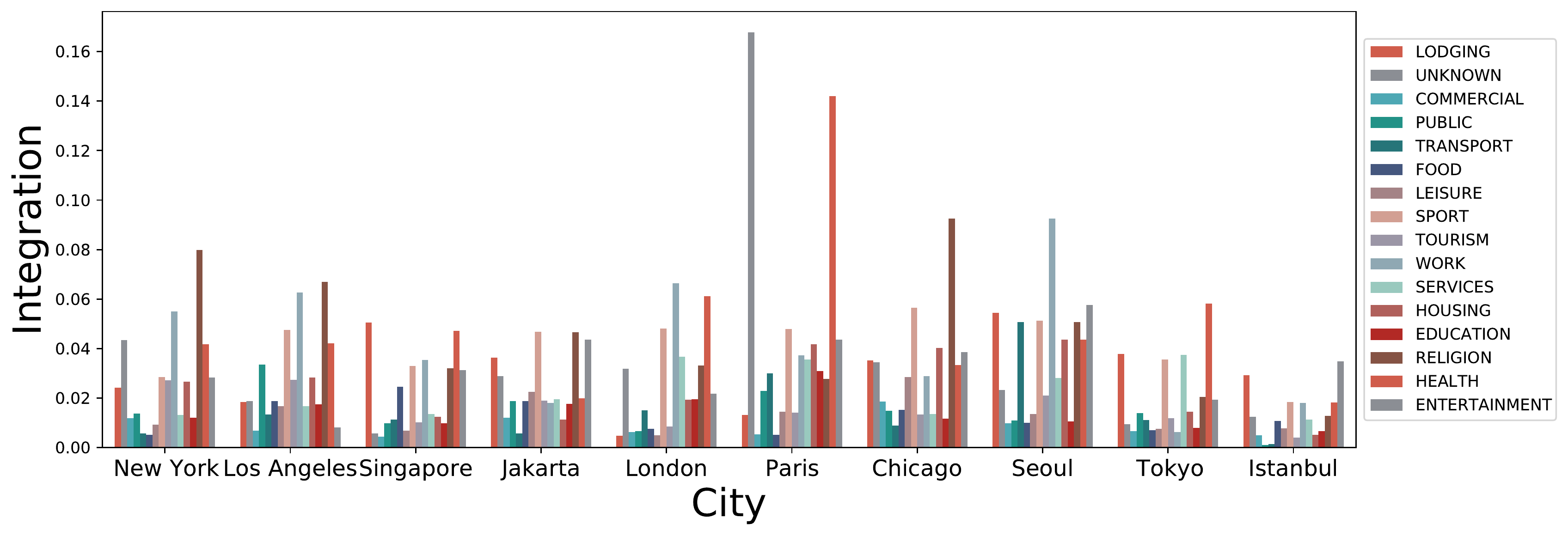} 
\includegraphics[width=\textwidth]{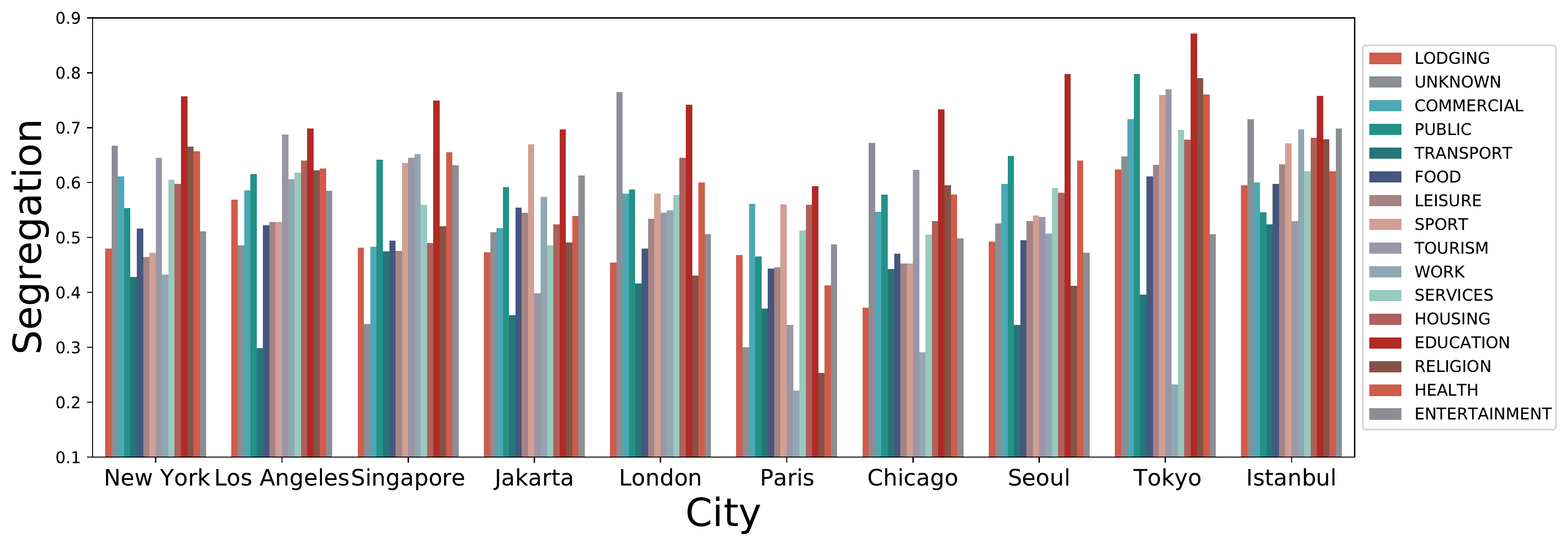} 
\caption{\textbf{Single Layer Segregation and Integration.}}
\label{SI:single_layer}
\end{figure*}

\begin{figure*}
\renewcommand{\figurename}{Supplementary Figure}
\begin{center}
\begin{tabular}{c}
\includegraphics[angle=0, width=0.70\textwidth]{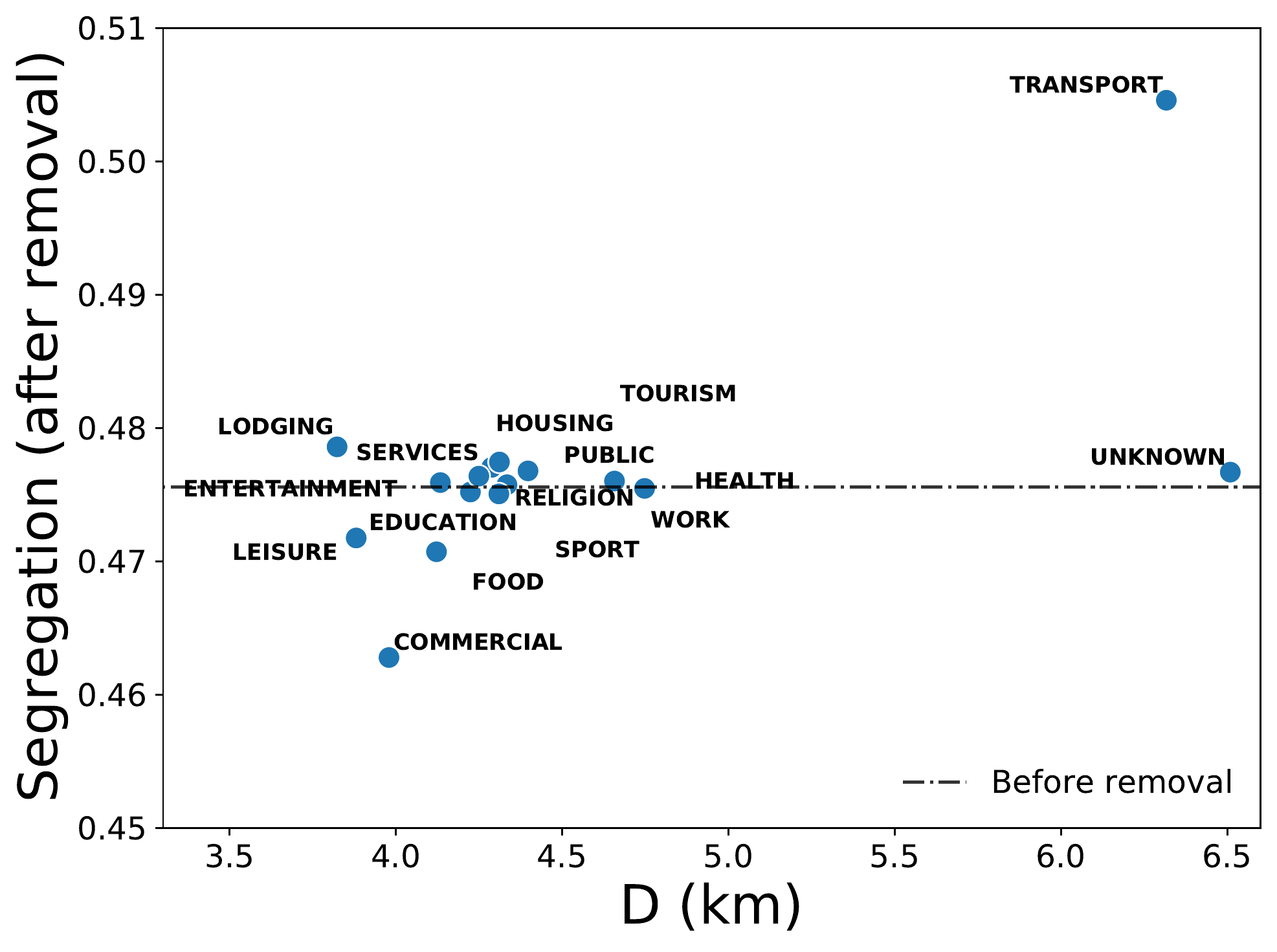}
\end{tabular}
\end{center}
\caption{
{\bf Average functional integration for different activity categories.} Conversely from what observed in Fig 5c for integration, we observe no clear dependency of the effect of removing a layer with  the average distance covered $D$ in movement inside that layer. Again, the transport layer is displaying exceptional behaviour.}
\label{SI:Q_removal_D}
\end{figure*}

\end{document}